\documentclass{elsart}
\usepackage{graphicx}
\usepackage{amssymb}
\usepackage{amsmath}

\begin{document}

\begin{frontmatter}

\title{Elaborating Transition Interface Sampling Methods}

\author[cecam]{Titus S. van Erp \corauthref{cor1}}
\corauth[cor1]{Corresponding author.}
\ead{tsvanerp@cecam.fr}
\author[UvA]{Peter G. Bolhuis}
\ead{bolhuis@science.uva.nl}
\address[cecam]{Laboratoire de Physique / Centre Europ\'een
de Calcul Atomique et Mol\'eculaire, Ecole Normale Sup\'erieure de
Lyon, 46 all\'ee d'Italie, 69364 Lyon Cedex 07, France}
\address[UvA]{Department of Chemical Engineering, Universiteit van Amsterdam, 
Nieuwe Achtergracht 166, 1018 WV Amsterdam, The Netherlands}

\begin{abstract}
We review two recently developed efficient methods for calculating rate
constants of processes dominated by rare events
in high-dimensional complex systems. The first is transition
interface sampling (TIS), based on the measurement of effective fluxes
through hypersurfaces in phase space. TIS improves efficiency with
respect to standard
transition path sampling (TPS) rate constant techniques,
because it allows a variable path length and is less sensitive
to recrossings. The second method is the  partial path version of TIS.
Developed for diffusive processes, it exploits the loss of long time
correlation.  We discuss the relation between the new techniques and
the standard reactive flux methods in detail.
Path sampling algorithms can suffer from ergodicity problems, and we
introduce
several new techniques to alleviate these problems, notably path swapping,
stochastic configurational bias Monte Carlo shooting moves and order-parameter
free path sampling.  In addition, we give algorithms to calculate other interesting
properties from path ensembles besides rate constants, such as  activation
energies and reaction mechanisms.
\end{abstract}

\end{frontmatter}

\section{Introduction}
Molecular simulation has become indispensable as a modern tool to gain
insight in the kinetics of processes in complex environment by supplying
detailed atomistic information that is not (easily) experimentally accessible.
Using either classical or {\it ab initio} based  atomistic force fields~\cite{Leach,CaPa85}, techniques such as molecular dynamics (MD)~\cite{AllenTildesley,FrenkelSmit}
can  model reactive events on a reasonable realistic level.
In contrast to most experiments 
where  kinetic properties such as the reaction
rate are obtained by measuring the macroscopic
population densities of reactant and product states over a long
time (seconds),  molecular dynamics simulations have to obtain good statistics
with much smaller systems (usually $\sim$ 100 to 100000 molecules) in
the accessible time range of nanoseconds-microseconds using a time
step of a few femtoseconds, as dictated by the molecular vibrations.
This small timescale and system size
limits the application to activated processes 
with relatively low barriers between reactant and product states. 
The computation of rate constants with straightforward MD becomes inefficient
when the process of interest has to overcome a high activation barrier
because the probability to observe a reactive event 
on this time- and system-scale
decreases exponentially with the barrier height. 
The system will spend a long time in one of the stable states and 
occasionally jump --in relatively short time--  to the other state.
This separation of  time scales results  in  two state kinetics: the exponential relaxation of the 
population densities~\cite{Chandlerbook}. 

The time-scale problem  is traditionally solved by a two-step reactive
 flux method~\cite{Keck62,Anderson73,Bennet77,DC78}. One first calculates
the free energy as a function of a reaction coordinate describing
the process. The transition state theory (TST) rate constant is then
related to the probability to be at the maximum of the free energy
barrier. This rate is only an approximation and the second part of the
reactive flux methods computes the correction, the transmission
coefficient, by starting many fleeting trajectories from the top of the
barrier~\cite{Keck62,Anderson73,Bennet77,DC78}.  However, the success
of this method depends strongly on the choice of reaction
coordinate. If the reaction coordinate fails to capture the molecular
mechanism the corresponding transmission coefficient will be extremely low,
making an accurate evaluation of the rate problematic if not
impossible.  For high dimensional complex systems, for instance chemical
reactions in solution, or protein folding, a good reaction coordinate
can be extremely difficult to find and usually requires detailed {\em
a priori} knowledge of the transition mechanism. Hence, TST based
reactive flux methods will be ineffective for complex processes for
which no prior knowledge is available.

Chandler and collaborators~\cite{TPS98,TPS98a,TPS99,Bolhuis02,Dellago02} 
devised a method for which no
reaction coordinate is needed, but only a definition of the reactant and product state.  This method, called transition path
sampling (TPS), gathers a collection of trajectories connecting the
reactant to the product stable region by employing a Monte Carlo (MC)
procedure called {\em shooting} and {\em shifting}. The resulting {\em
path ensemble} gives an
unbiased insight in the mechanism of the reaction.  TPS
has been successfully applied to such diverse systems as cluster
isomerization, auto-dissociation of water, ion pair dissociation and
on isomerization of a dipeptide, as well a reactions in aqueous
solution (see Ref.~\cite{Bolhuis02} for an overview). 
A drawback of TPS is that the calculation of rate constants is rather computer time consuming.  We
therefore developed the more efficient transition interface sampling
(TIS) method \cite{ErpMoBol2003}.  TIS allows a variable path length, thereby
limiting the required MD time steps to the strict necessary
minimum.  The TIS rate equation is based on an
effective positive flux formalism and is less sensitive to
recrossings.  The shifting moves used in TPS to enhance
 statistics, are unnecessary in the TIS algorithm.  Also,
multidimensional or even discrete order parameters can easily be
implemented in TIS. 
 Recently, we showed that for diffusive processes
one can exploit the loss of correlation along trajectories. This lead
to the development of the partial path TIS (PPTIS) method, a
variation of TIS that samples much shorter paths~\cite{MoBolErp2004}.

In this paper we re-derive the basic concepts of TIS and PPTIS in a
more intuitive way and relate them 
to the calculation of the transmission coefficient.
For the mathematical validation of the expressions
we refer to Refs. \cite{ErpMoBol2003,MoBolErp2004}.  
The paper is organized as follows. In section II we discuss the relation 
between several different microscopic expressions for the phenomenological 
rate constant present in the literature, and derive the positive effective 
flux formalism on which both interface path sampling methods are based. In section III we 
present the TIS and PPTIS formalism and  precise algorithm.  In section IV, 
we introduce new algorithms for alleviating 
ergodicity problems that might occur in path sampling simulations. The last 
section V, is reserved for new ways of extracting interesting properties  from path ensembles, such as  the activation energy of a reaction. We end 
with concluding remarks.

\section{Microscopic rate equations}
The calculation of reaction rate constants by computer simulation
requires an expression for the rate constant in terms of microscopic
properties. 
Such a microscopic rate expression needs a proper characterization of the 
reactant state $A$ and product state $B$ for each
separate reaction, but   
should not be too sensitive to these state definitions, otherwise an unrealistic ill-defined rate will result.
Once we have a rate expression, there are  several  ways to compute
the reaction rate. The standard reactive flux method measures the flux 
through a single hypersurface in phase-space dividing the
reactant state $A$ from the product state $B$. In TPS the rate constant is
taken from a time derivative of a correlation function, which can be
calculated by slowly confining a completely free path ensemble to an
ensemble that connects reactant to product. 
 The TIS approach measures a reactive
flux 
through
many interfaces between $A$ and $B$.  
These three methods can be related to each other, as they
ultimately compute the same properties. The TPS 
correlation function at $t=0$
becomes equivalent to the
TST approach when $A$ and $B$ are adjacent in phase space \cite{Dellago02}.
The TIS
effective positive flux formalism for a single interface is equivalent
to TST-based transmission coefficient calculations\cite{ErpMoBol2003}. The TIS rate
equation can also be recast in terms of a TPS-like correlation
function but then based on the so-called \emph{overall} states of the system. 
In the following subsections, we will explain the reactive flux, 
TPS, and TIS methods and their connections in detail.

\subsection{Transition state theory}

The first step in TST is to choose a reaction coordinate $\lambda$
 describing the transition from a stable reactant state $A$ to a stable 
product state $B$. This reaction coordinate can be any function $\lambda(x)$ 
of phase space point $x\equiv \{r,p\}$, with $r$ the particle coordinates and 
$p$ the momenta. Next, the free energy  $F(\lambda)=-k_B T \ln(P(\lambda))$ is 
calculated by determining the probability $P(\lambda)$ to be at $\lambda$ 
using, for instance,
biased sampling techniques~\cite{TV74,CCH89,Ciccotti91}.  Here, $k_{B}$ is the 
Boltzmann constant and $T$ is the temperature. 
The maximum $\lambda^*$ in $F(\lambda)$ defines the dividing surface
$\{x|\lambda(x)= \lambda^*\}$ separating state $A$ from state $B$.
By convention, the system is in $A$ if $\lambda(x)<\lambda^*$ and in
$B$ if $\lambda(x)>\lambda^*$.  For a phase point $x$ in $A$, the
probability to be at the top of the barrier is:
\begin{eqnarray}
\label{FreeEn}
P(\lambda^*)_{x \in A} \equiv \frac{\left \langle
\delta(\lambda(x)-\lambda^*) \right \rangle }{\left \langle
\theta\big( \lambda^* - \lambda(x)\big) \right
\rangle}=\frac{e^{-\beta F(\lambda^*)}}{\int_{-\infty}^{\lambda^*}
{\mathrm d} \lambda \, e^{-\beta F(\lambda )} },
\end{eqnarray}
where the brackets $\left \langle \ldots \right \rangle$ denote the
equilibrium ensemble averages, $\theta(x)$ and $\delta(x)$ are the
Heaviside step-function and the Dirac delta function, respectively,
and $\beta=(k_{B} T)^{-1}$.  
TST  assumes that trajectories that cross $\lambda^*$ do
not recross the dividing surface 
Hence, the TST
expression is equivalent to the positive flux through the dividing surface
$\lambda^*$:
\begin{eqnarray}
\label{rateTST}
k_{AB}^{TST}= \left \langle \dot{\lambda}(x) \theta\big(
\dot{\lambda}(x) \big) \right \rangle_{\lambda^*} P(\lambda^*)_{x \in
A},
\end{eqnarray}
where the dot denotes a time derivative and the subscript $\lambda^*$ to the ensemble brackets indicates
that the ensemble is constrained to the top of the barrier on the
dividing surface $\lambda^*$.  The TST rate constant is 
sensitive to  the choice of reaction coordinate
$\lambda(x)$ and will only be correct  if the
surface $\{x|\lambda(x)=\lambda^*\}$ corresponds to the true
transition state dividing surface: the so-called separatrix.  
For complex systems, it is impossible to know the location and shape
of this curved multidimensional separatrix.  It is possible, however,
to correct the TST expression with a dynamical factor  that is called the
transmission coefficient.

\subsection{Transmission coefficients}
\label{sectrans}
Traditionally, the dynamical corrected rate constant is derived by
applying a small perturbation to the equilibrium state and invoking
the fluctuation-dissipation theorem \cite{Yamamoto60,Chandlerbook,FrenkelSmit}. 
This leads,  for instance, to  the well known Bennett-Chandler (BC) \cite{Bennet77,DC78}
expression for the reaction rate 
\begin{eqnarray}
\label{BCrate}
k_{AB}^{BC}(t)= \frac{\left \langle \dot{\lambda}(x_0) \delta\big(
\lambda(x_0)-\lambda^*\big) \theta\big( \lambda(x_t)-\lambda^* \big)
\right \rangle}{ \left \langle \theta\big( \lambda^* - \lambda(x_0)
\big) \right \rangle}
\end{eqnarray}
where $x_t$ specifies the  coordinates and momenta of the system
at time $t$ as obtained from  a short molecular dynamics (MD) trajectory starting at $x_0$.
The ensemble average is taken over all phase points $x_0$.
For exponentially relaxing two state kinetics with a well defined  rate 
constant,  
there is a separation of timescales:
 the reaction time
$\tau_{rxn}$ (or  expectation time for one single event) is much
longer than the molecular time $\tau_{mol}$ that the system spends on
the barrier. In that case, Eq.~(\ref{BCrate}) 
will reach a plateau value for  $\tau_{mol} \ll t \ll \tau_{rxn}$, which is equal to the correct  phenomenological rate constant $k_{AB}$.
The function  $k_{AB}^{BC}(t)$ will
sensitively depend on the choice of the reaction coordinate $\lambda$, 
but the plateau value
will not.  In the limit $t \rightarrow 0^+$,  the  BC rate reduces to the TST expression Eq.~(\ref{rateTST}) 

\begin{figure}[t]
\begin{center}
\includegraphics[width=4cm,keepaspectratio]{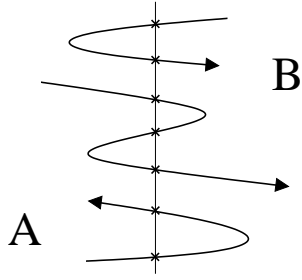}
\end{center}
\caption{Illustration of the difference in counting in the transmission 
coefficient Eqs.~(\ref{BCtrans}), (\ref{transBC2}), and (\ref{transTIS}).
For simplicity, assume that the system consists of three kind of possible 
trajectories, as shown by this figure,  that cross the dividing 
surface with the same speed $v$ orthogonal to the surface.
To the seven phase points on the surface (from top to bottom)
the numerator of Eq.~(\ref{BCtrans}) will assign the 
values $[-v,v,v,-v,v,0,0]$, while these are $[0,0,v,-v,v,0,0]$ for 
Eq.~(\ref{transBC2}) and $[0,0,v,0,0,0,0]$ for Eq.~(\ref{transTIS}). 
The sum of these give the same result $v$. 
Evaluation of Eq.~(\ref{transTIS}) in an actual computer algorithm 
requires the fewest MD steps as only  phase points similar to the 
3rd and 7th  phase points
would need the integration until reaching stable state regions.
For instance,  
the fifth crossing point can be assigned zero already as soon 
as one detects that its backward trajectory recrosses the surface.}
\label{fluxcancel}
\end{figure}

The transmission coefficient is defined as the ratio
between the real rate constant and the TST expression: $\kappa \equiv k_{AB}/
k_{AB}^{TST}$:
\begin{eqnarray}
\label{BCtrans}
\kappa^{BC}(t)= \frac{\left \langle \dot{\lambda}(x_0) \theta\big(
\lambda(x_t)-\lambda^* \big) \right \rangle_{\lambda^*}}{ \left
\langle \dot{\lambda}(x_0) \theta\big( \dot{\lambda}(x_0) \big) \right
\rangle_{\lambda^*} }
\end{eqnarray}
The numerator in Eq.~(\ref{BCtrans})  counts trajectories
with a positive but also with a negative weight. The latter 
trajectories  leave  the surface at $t=0$ with a negative
velocity $\dot{\lambda}(x_0)$, but are eventually found at the B
side of the surface after a (few) recrossing(s).  
However, 
untrue $B \rightarrow B$ trajectories do not contribute to the rate because the
 positive and negative terms 
cancel~\footnote{This
cancellation might seem to be not apparent if a trajectory recrosses
the same surface but with a different velocity. Still, this is the
case. The absolute value of the flux of a trajectory is at each
intersecting surface the same.  A lower crossing velocity
$\dot{\lambda}$ is compensated by a higher probability to measure the
crossing point as the trajectory spends more time at the surface.}
(See Fig.~\ref{fluxcancel}).  
Similarly, the
$A\rightarrow B$ trajectories with multiple $\lambda^*$ crossings are
effectively counted only once~\cite{ErpMoBol2003}.  Although Eq.~(\ref{BCrate})
gives the correct rate constant, it is rather
unsatisfactory to sample 
only trajectories forward in time not knowing which  contribute to the rate and which do not. 
Therefore, alternative expressions for the rate constant have been 
proposed taking the 
past  into account.
Here, they  are  referred to as the BC2
\cite{Bennet77,DC78} expression
\begin{eqnarray}
\label{transBC2}
\kappa^{BC2}(t) &=& \frac{\left \langle \dot{\lambda}(x_0)
\theta\big(\lambda^*- \lambda(x_{-t}) \big) \theta\big(
\lambda(x_t)-\lambda^* \big) \right \rangle_{\lambda^*}}{ \left
\langle \dot{\lambda}(x_0) \theta\big( \dot{\lambda}(x_0) \big) \right
\rangle_{\lambda^*} }
\end{eqnarray}
and the positive flux PF \cite{Bergsma86} expression
\begin{eqnarray}
\label{transpf}
\kappa^{pf}(t) &=& \frac{\left \langle \dot{\lambda}(x_0) \theta\big(
 \dot{\lambda}(x_0) \big) \theta\big( \lambda(x_t)-\lambda^* \big)
 \right \rangle_{\lambda^*}}{ \left \langle \dot{\lambda}(x_0)
 \theta\big( \dot{\lambda}(x_0) \big) \right \rangle_{\lambda^*} }
 - \frac{\left \langle \dot{\lambda}(x_0) \theta\big(
 \dot{\lambda}(x_0) \big) \theta\big( \lambda(x_{-t})-\lambda^* \big)
 \right \rangle_{\lambda^*}}{ \left \langle \dot{\lambda}(x_0)
 \theta\big( \dot{\lambda}(x_0) \big) \right \rangle_{\lambda^*} }\nonumber \\
\end{eqnarray}
In Eq.~(\ref{transBC2}) the theta functions guarantee 
that only true  $A \rightarrow B$ events are counted. Still, the numerator in 
Eq.~(\ref{transBC2})
contains  negative terms: those phase points $x_0$ with a negative velocity
$\dot{\lambda}(x_0)$ and with corresponding backward and forward
trajectory that ends up in $A$ and $B$, respectively.
Eq.~(\ref{transpf}) counts only positive crossings,
but cancellation with a negative term can occur when the backward
trajectory also ends  up at the $B$ side of the dividing surface.  At
first sight, Eq.~(\ref{transpf}) seems  to
overcount  $A\rightarrow B$ trajectories with multiple
$\lambda^*$ crossings.  However, if one realizes that each $A \rightarrow B$ 
trajectory has an equivalent trajectory $B \rightarrow A$ by reversing the 
time, an overall cancellation of positive and negative terms ensures a proper 
final outcome.

For completeness, we mention that there are also similar expressions
by Berne \cite{Berne85,White2000} and a relation by Hummer
\cite{Hummer04} that counts both positive and negative crossings with
a positive weight, but only if the corresponding trajectory ends at
opposite sides of the surface and with a weight lower than 
$|\dot{\lambda}|$ if its trajectory has more than just one crossing.
Ruiz-Montero et al. designed a transition
zone method in which they measure the flux on many places at the top
of the barrier and weight them to the inverse free energy
\cite{Ruiz97}.

\subsection{The effective positive flux formalism}

A more intuitive, yet sound, alternative to the above expressions
is the \emph{ effective} flux formalism.  
We can illustrate this formalism with an analogy to the
migration of people from country $A$ to $B$.  
To determine the emigration rate we can  
simply count the number of persons that cross the border from $A$ to
$B$ within a certain time interval.  
However, we should not count
tourists. This group consist of people who have a nationality $A$ and
will only spend a short time in $B$, or have a nationality $B$ and are
actually on their way back.  Moreover, we have to be aware that some
emigrants might cross the frontier several times on their way. To
prevent overcounting, we should only count one specified crossing for each person, for instance, the first or the last crossing of the 
emigration journeys from $A$ to
$B$.  The same reasoning can be applied when calculating the
rate constant of a reaction.  In a molecular simulation we can check
the 'nationality' of the system and the one-crossing condition by
simply following the equations of motions backward and forward in
time.  This procedure, to count only true events and to avoid counting
recrossings is what we call the {\it effective positive flux} formalism.  In
Sec. \ref{subsecCWT} we give the mathematical expression of the
effective positive flux.

It is surprising that the effective
positive flux  counting strategy is not so common.  To our knowledge only two slightly different
expressions of a transmission coefficient based on the effective
positive flux have been proposed in
Refs.~\cite{Anderson95,titusthesis}.  
In all other expressions found in the literature the
counting of recrossings is not avoided, but the final rate
constant follows through cancellation of many negative and positive
terms.  The effective flux transmission coefficients formulation 
is most useful  when applying  a single dividing surface and when
recrossings are apparent~\cite{White2000}. In general, we note that 
any averaging
method  counting only zero and positive values will show a faster
convergence than  one that is based on  cancellation of positive
en negative terms. Moreover, in the effective flux formalism many
trajectories will be assigned as unreactive after just a few MD steps 
(See Fig.~\ref{fluxcancel}), thus reducing the number of required force 
evaluations.  
A comparative
study of ion channel diffusion \cite{White2000} showed that the
algorithm based on effective positive flux expression of Anderson
\cite{Anderson95} was superior to the other transmission rate
expressions. Moreover, it was found as efficient as an optimized
version of the more complicated Ruiz-Montero method \cite{Ruiz97}.

\subsection{TPS correlation function}
\label{sectpscorr}
In TPS one also has to define an order parameter $\lambda(x)$, but
this does not have to be a properly chosen reaction coordinate
capturing the essence of the dynamical mechanism. Instead, it is
sufficient but necessary 
that this function is able to characterize the basins of
attraction of the stable states~\cite{Bolhuis02}.  By definition
the system is in $A$ if $\lambda(x)< \lambda_A$ and in $B$ if
$\lambda(x) > \lambda_B$ with $\lambda_A < \lambda_B$.  Clearly, 
the two states are not connected and the intermediate
 barrier region, belongs neither to $A$ nor to $B$.  By
introducing following characteristic functions
\begin{eqnarray}
h_A(x) &=& 1, \quad \textrm{ if } x \in A, \quad \textrm{ else } \quad
h_A(x)=0 \nonumber \\ h_B(x) &=&1, \quad \textrm{ if } x \in B, \quad
\textrm{ else } \quad h_B(x)=0.
\label{charAB}
\end{eqnarray}
the TPS-correlation function is defined as:
\begin{equation} 
C(t)= \frac{ \left \langle h_A(x_0) h_B(x_t) \right \rangle }{\left
\langle h_A(x_0) \right \rangle }.
\label{corrTPS}
\end{equation}
If there is a  
separation of timescales, this correlation
function grows linearly in time, $C(t) \sim k_{AB} t$, for times $\tau_{mol} <t <\tau_{rxn}$
In that case,  the time
dependent reaction rate
\begin{equation} k_{AB}^{TPS}(t)=
\frac{\mathrm{d}}{\mathrm{d}t} C(t)
\label{rateTPS}
\end{equation}
reaches a plateau for
$\tau_{mol} < t < \tau_{rxn}$.  $C(t)$
can be calculated in a path sampling simulation employing the 
shooting and shifting Monte Carlo moves, in combination with  
an umbrella sampling algorithm in which the final region $B$
is slowly shrunk from the entire phase space  to the final stable state 
$B$~\cite{Dellago02}.  The disadvantage of such a procedure 
is that it can take a relatively long
time $\tau_{mol}$ before $C(t)$ reaches a plateau 
(longer than in a transmission coefficient calculation \cite{Dellago02}). 
 
All paths in the path sampling should have a minimal length 
$\mathcal{T}>\tau_{mol}$ 
and as a result unnecessarily long periods are spent inside the stable state 
basins of attraction.  Moreover, inspection of Eqs.~(\ref{corrTPS}) and 
(\ref{rateTPS}) shows that a necessary 
cancellation of positive and negative terms  
can slow down the convergence of the MC sampling procedure.
In the case of adjacent $A$ and $B$ regions, the TPS formalism becomes 
equivalent to the TST approximation in the limit 
$t\rightarrow 0$ \cite{Dellago02}.

\subsection{The road to TIS}
\label{subsecCWT}
The TIS method is based on the measurement of the fluxes though
multiple dividing surfaces. Consider a set of $n+1$
non-intersecting multidimensional interfaces $\{0,1 \dots n \}$
described by an order parameter $\lambda(x)$ that does
not have to correspond to the real reaction coordinate.  We choose
$\lambda_i$, $i=0\dots n$ such that $\lambda_{i-1} < \lambda_i$, and
that the boundaries of state A and B are described by $\lambda_0$ and
$\lambda_n$, respectively.  For each phase point $x$ and each
interface $i$, we define a backward time $t_i^b(x)$ and forward time
$t_i^f(x)$:
\begin{eqnarray}
t_i^b(x_0) &\equiv& -\max \left[ \{ t | \lambda(x_t) = \lambda_i
\wedge t \leq 0 \} \right] \nonumber \\ t_i^f(x_0) &\equiv& +\min
\left[ \{ t | \lambda(x_t) = \lambda_i \wedge t \ge 0 \} \right],
\end{eqnarray}
which mark the points of first crossing with interface $i$ on a
backward (forward) trajectory starting in $x_0$.  Note that $t_i^b$
and $t_i^f$ defined in this way always have positive values. Following
Ref.~\cite{ErpMoBol2003}, we then introduce two-fold
characteristic functions that depend on two interfaces $i \neq
\nobreak j$,
\begin{eqnarray}
\bar{h}_{i,j}^b(x)  = 
\begin{cases}
1 \quad \textrm{ if } t_i^b(x) < t_j^b(x) ,\\ 0 \quad \textrm{
otherwise}
\end{cases}, 
\qquad
\bar{h}_{i,j}^f(x)  = 
\begin{cases}
1 \quad \textrm{ if } t_i^f(x) < t_j^f(x) ,\\ 0 \quad \textrm{
otherwise}
\end{cases} \
\end{eqnarray}
which measure whether the backward (forward) time evolution of $x$
will reach interface $i$ before $j$ or not.  However, as the
interfaces do not intersect, the time evolution has to be evaluated
only for those phase points $x$ that are in between the two interfaces
$i$ and $j$. In case $i<j$, we know in advance that $t_i^{b,f}(x) <
t_j^{b,f}(x)$ if $\lambda(x)<\lambda_i$ and $t_i^{b,f}(x) >
t_j^{b,f}(x)$ if $\lambda(x)>\lambda_j$.  When the system is ergodic,
both interfaces $i$ and $j$ will be crossed in finite time and thus
$\bar{h}_{i,j}^b(x) + \bar{h}_{j,i}^b(x) = \bar{h}_{i,j}^f(x) +
\bar{h}_{j,i}^f(x)=1$.  The two backward characteristic functions
define the TIS {\it overall} states ${\mathcal A}$ and ${\mathcal B}$:
\begin{eqnarray}\label{TISstates}
h_{\mathcal A}(x)=\bar{h}_{0,n}^b(x), \quad h_{\mathcal
B}(x)=\bar{h}_{n,0}^b(x).
\end{eqnarray}
Together, the {\it overall} states cover the entire phase space and,
within certain limits, do not sensitively depend on the precise
boundaries of stable states $A$ and $B$.  With these new
characteristic functions  
we can write down a
correlation function similar to Eq.~(\ref{corrTPS}):
\begin{equation}
C(t)= \frac{\left \langle h_{\mathcal A}( x_0 ) h_{\mathcal B}( x_t )
\right \rangle }{ \left \langle h_{\mathcal A}(x_0) \right \rangle},
\label{corrTIS}
\end{equation}
This correlation function exhibits a linear regime $\sim k_{AB}
t$ for $0 < t< \tau_{rxn}$. Therefore, we can simply take the time
derivative at $t=0$ yielding
\begin{eqnarray} \label{rateTIS}
k_{AB}= \frac{ \left \langle \bar{h}_{0,n}^b(x_0) \dot{\lambda}(x_0)
\delta( \lambda(x_0)-\lambda_n) \right \rangle }{ \left \langle
h_{\mathcal A}(x_0) \right \rangle }.
\end{eqnarray}
One can easily verify that here only positive terms contribute to the rate.
The connection to the transmission coefficient can be made
by using following relation~\cite{ErpMoBol2003}:
\begin{eqnarray}
\label{fluxrel}
 \left \langle \bar{h}_{i,k}^b \dot{\lambda}\delta(
 \lambda(x)-\lambda_k) \right \rangle =
 \left \langle \bar{h}_{i,j}^b \dot{\lambda}\delta(
 \lambda(x)-\lambda_j) \bar{h}_{k,i}^f \right \rangle
\end{eqnarray}
for $\lambda_i < \lambda_j < \lambda_k$.  Using this equality, we can
write down a transmission coefficient similar to the ones in
Sec. \ref{sectrans}
but then based on the effective positive flux \cite{titusthesis}:
\begin{eqnarray}
\label{transTIS}
\kappa^{TIS}= \frac{\left \langle \bar{h}_{0,i}^b(x_0)
\dot{\lambda}(x_0) \theta\big( \dot{\lambda}(x_0) \big)
\bar{h}_{n,0}^f(x_0) \right \rangle_{\lambda_i}}{ \left \langle
\dot{\lambda}(x_0) \theta\big( \dot{\lambda}(x_0) \big) \right
\rangle_{\lambda_i} }
\end{eqnarray}
for $\lambda_i=\lambda^*$.  Although, in principle $ \theta\big(
\dot{\lambda}(x_0) \big)$ is redundant in the numerator of
Eq.~(\ref{transTIS}) as $\bar{h}_{0,i}^b(x_0)=0$ if
$\dot{\lambda}(x_0)<0$,  it is there to highlight that only
positive crossings are counted. Trajectories started at $x_0$ on interface 
$i$ are followed backward 
in time  until they reach stable region $A$ or recross interface $i$.
Then, only the ones that reach stable region $A$ are also followed 
forward in time until they reach one of the stable regions. 
The slightly different effective flux expression
of Ref.~\cite{Anderson95} follows trajectories until 
reaching the plateau region time and counts
for each $A\rightarrow B$ trajectory only the last crossing instead of the 
first.

\section{(Partial path) transition interface sampling}
\subsection{Formalism}
In a system for which the correct
reaction coordinate $\lambda$ is known in advance 
and that is not 
dominated by recrossings, the effective positive flux formalism
of (Eq.~(\ref{transTIS}) and Ref.~\cite{Anderson95}) is
probably the best choice when using a single dividing
surface \cite{White2000}. 
However, for complex systems, for instance chemical
reactions in solution, any intuitively chosen reaction coordinate can
give arbitrary small transmission coefficients, making an accurate
computation prohibitive.  To improve reaction coordinates by
e. g. taking solvent degrees into account is generally a difficult job.
Some progress has been made by using the coordination number as
reaction coordinate \cite{Sprik98,Sprik2000}, but this ad hoc approach  probably
only works for specific systems. For instance, we showed that a
proton transfer reaction in water depends very sensitively on the
angular orientation of the surrounding water molecules \cite{ErpMeij04}.  
Similarly, the degrees of freedom in a protein are so large that dynamical 
folding processes are at best only very qualitatively described by order 
parameters. Quantities such as radius of gyration or number of native contact 
do usually not correspond to reaction coordinates \cite{BolhuisPNAS}.
Subtle effects, e.g. the solvent structure, play also here a role.
To incorporate all these subtleties in a
single one-dimensional reaction coordinate is an immense task
and can only be successful if the precise reaction mechanism is already
known in advance.  The TPS and TIS techniques do not rely on a reaction
coordinate. The TIS hypersurfaces  do not have to coincide with  the transition 
state dividing surface.  

At the end of this section we give TIS (and PPTIS) rate 
expressions that can be employed in a computer algorithm.  First, as the 
derivation of the TIS and PPTIS formalism requires a proper notation,  we 
introduce following flux function
\begin{eqnarray} \label{fluxij}
\phi_{ij}(x) \equiv  \bar{h}_{j,i}^b(x) |\dot{\lambda}(x)| \delta(
\lambda(x)-\lambda_i) 
= \bar{h}_{j,i}^b(x) \lim_{\Delta t \rightarrow 0} \frac{1}{\Delta
t} \theta\big( \Delta t- t_i^f(x) \big)
\end{eqnarray}
The first equality  has the same flux notation as
Eq.~(\ref{rateTIS}), but the second equality is more useful
in practice.  An MD trajectory might cross interface
$\lambda_i$, but consists of discrete time slices that are never
exactly on the surface  (as opposed to a transmission
coefficient calculation).  However, $\phi_{ij}(x)$ can still be defined for the
discrete MD set of time-slices by taking $\Delta t$ equal to the
molecular time-step.  In words, $\phi_{ij}(x)$  equals 
$1/\Delta t$ if the forward trajectory crosses $\lambda_i$ in one
single $\Delta t$ time-step and the backward
trajectory crosses $\lambda_j$ before $\lambda_i$.  Otherwise $\phi_{ij}(x)$ vanishes.  In addition, we introduce a flux function that incorporates
also the forward trajectory
\begin{eqnarray}
\Phi_{ij}^{lm} (x) \equiv \phi_{ij}(x) \bar{h}_{l,m}^f(x)
\end{eqnarray}
By making use of Eq.~(\ref{fluxrel}) we can write for $\lambda_i < \lambda_j <
\lambda_k$:
\begin{eqnarray}
\left \langle \phi_{ki}(x) \right \rangle = \left \langle
\Phi_{ji}^{ki} (x) \right \rangle
\end{eqnarray}
and, thus, the rate constant (\ref{rateTIS}) becomes
\begin{equation}
\label{eq:kab_phi}
k_{AB}= \left \langle \phi_{n,0} \right \rangle / \left \langle
h_{\mathcal A} \right \rangle = \left \langle \Phi_{i,0}^{n,0} \right
\rangle / \left \langle h_{\mathcal A} \right \rangle 
\end{equation}
for each $\lambda_i$ with $ 0 \leq i \leq n$. 

\begin{figure}[b]
\begin{center}
\includegraphics[width=6cm,keepaspectratio]{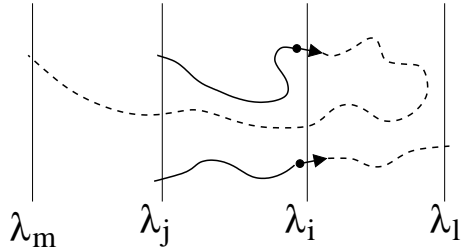}
\end{center}
\caption{The conditional crossing probability $P(_m^l|_j^i)$  
for a certain configuration of interfaces 
$\lambda_i, \lambda_j, \lambda_l$, and $\lambda_m$. The condition $|_j^i)$ is
depicted by the arrow and the solid line for two phase points (the dots):
from this phase point one should cross $\lambda_i$ in one single $\Delta t$
time-step in the forward direction, and, besides, 
its backward trajectory should cross  $\lambda_j$
before $\lambda_i$. Two possible forward trajectories are given by the dashed 
line. The upper crosses $\lambda_m$ before $\lambda_l$, the lower
crosses  $\lambda_l$ as first. The fraction 
whose forward trajectories behave like the last case equals
$P(_m^l|_j^i)$.}
\label{figprob}
\end{figure}

The second step is to
define a conditional crossing probability that depends on the
location of four interfaces:
\begin{eqnarray}
\label{P4int}
P(_m^l|_j^i) \equiv {\left< \Phi_{ij}^{lm} \right>}/ {\left< \phi_{ij}
\right>}.
\end{eqnarray}
In words, this is the probability for the system to reach interface
$l$ before $m$ under the condition that it crosses at $t=0$ interface
$i$, while coming directly from interface $j$ in the past (see
Fig.~\ref{figprob}).
The probabilities in Eq.~(\ref{P4int}) are the building blocks for both TIS 
as PPTIS to construct expressions for the rate constant.  The
probabilities $P(_m^l|_j^i)$ are defined on any set of four 
interfaces. The case, where  $m=j=0$ and $m=j=n$, is of special interest for
TIS and will be annotated as follows
\begin{equation}
\mathcal{P}_{A} (\lambda_j | \lambda_i) \equiv P(_0^j|_0^i), \quad
\mathcal{P}_{B} (\lambda_j | \lambda_i) \equiv P(_n^j|_n^i)
\end{equation}
For PPTIS, two types of  crossing probabilities  are required: the one 
interface crossing probabilities
\begin{eqnarray}
p_i^\pm & \equiv & P(_{i-1}^{i+1}|_{i-1}^i) , \qquad p_i^\mp \equiv
P(_{i+1}^{i-1}|_{i+1}^i) \nonumber \\ p_i^= & \equiv &
P(_{i+1}^{i-1}|_{i-1}^i), \qquad p_i^\ddagger \equiv
P(_{i-1}^{i+1}|_{i+1}^i),
\label{onehop}
\end{eqnarray}
and the long distance crossing probabilities
\begin{eqnarray}
P_i^+ \equiv P(_0^i|_0^1), \qquad P_i^- \equiv P(_{i}^0|_{i}^{i-1}).
\label{longhop}
\end{eqnarray}
Using these probabilities, the TIS rate constant can  be written
in terms that can be determined in a computer simulation~\cite{ErpMoBol2003}
\begin{eqnarray}
\label{kabphiP}
k_{AB} &=&\frac{\left \langle \phi_{1,0} \right \rangle}{ \left
\langle h_{\mathcal A} \right \rangle } {\mathcal
P}_A(\lambda_{n}|\lambda_1) \nonumber \\ {\mathcal
P}_A(\lambda_{n}|\lambda_1) &=& \prod_{i=1}^{n-1} {\mathcal
P}_A(\lambda_{i+1}|\lambda_i)
\end{eqnarray}
The first factor $\frac{\left \langle \phi_{1,0} \right \rangle}{ \left
\langle h_{\mathcal A} \right \rangle }$ is a flux and can be
calculated by straightforward MD
as $\lambda_1$ will be close to $A$ (see  Sec. \ref{secflux}).  The second factor, the crossing
probability ${\mathcal P}_A(\lambda_{n}|\lambda_1)$, is calculated
using the factorization in Eq.~(\ref{kabphiP}) into probabilities
${\mathcal P}_A(\lambda_{i+1}|\lambda_i)$ that are much higher than
the overall crossing probability.  These can be calculated using the
shooting algorithm as will be explained in Sec.~\ref{secshooting}. 

For PPTIS the
set of equations are as  follows\cite{MoBolErp2004}:
\begin{eqnarray}\label{ratePPTIS}
k_{AB} =& \frac{ \left \langle \phi_{1,0} \right \rangle}{ \left
\langle h_{\mathcal{A}}\right \rangle} P_n^+, 
\qquad k_{BA}&=\frac{ \left \langle \phi_{n-1,n} \right \rangle}{ \left \langle
h_{\mathcal{B}}\right \rangle} P_{n}^-  \\ 
P_j^+ \approx& \frac{p_{j-1}^\pm P_{j-1}^+}{p_{j-1}^\pm+p_{j-1}^= P_{j-1}^-}  
\label{recPPTIS}, \qquad   
P_j^- &\approx \frac{p_{j-1}^\mp
P_{j-1}^-}{p_{j-1}^\pm+p_{j-1}^= P_{j-1}^-}
\end{eqnarray}
The factor $\frac{\left \langle \phi_{1,0} \right \rangle}{ \left
\langle h_{\mathcal A} \right \rangle }$ is identical to the TIS flux factor, whereas to obtain the reverse rate $k_{BA}$ only a single extra factor
$\frac{ \left \langle \phi_{n-1,n} \right \rangle}{ \left \langle
h_{\mathcal B}\right \rangle }$ is needed.  The $P_n^+$ and $P_{n}^-$
are obtained via the recursive relations~(\ref{recPPTIS}) once all single 
crossing probabilities of Eq.~(\ref{onehop}) are known.  Starting with
$P_1^+=P_{1}^-=1$, we can iteratively determine $(P_j^+,P_j^-)$ for
$j=2,\ldots$ until $j=n$.  The one-hopping
probabilities~(\ref{onehop}) can again be  calculated using the shooting algorithm.
The PPTIS formalism basically transforms the process of interest
into a Markovian sequence of hopping events. Yet, if the dynamics is
diffusive and the interfaces are sufficient far apart, the rate
formalism (\ref{ratePPTIS}) and (\ref{recPPTIS}) will be almost exact 
\cite{MoBolErp2004}.

\subsection{The flux algorithm}
\label{secflux}
The flux factor $\frac{\left \langle \phi_{1,0}
 \right \rangle}{ \left \langle h_{\mathcal A} \right \rangle }$
 is the   effective flux through
 $\lambda_1$ of the trajectories coming from $\lambda_0$ (from
 $A$). This factor  is most conveniently computed with the first two 
interfaces identical. Although $\frac{\left \langle \phi_{1,0} \right \rangle}{
 \left \langle h_{\mathcal A} \right \rangle }$ is not well defined
 for $\lambda_1=\lambda_0$, we can 
think that
$\lambda_1=\lambda_0+\epsilon$ in the limit $\epsilon \rightarrow 0$.
In this way, the effective positive flux will be equal to the simple
positive flux through $\lambda_1$ 
(trajectories cannot recross without re-entering $A$, hence, all 
crossings are counted.).  
Similarly, for the reverse rate $k_{BA}$
we can set
 $\lambda_{n-1}=\lambda_n-\epsilon$.  If $\lambda_1$ is chosen 
 close enough to $A$ the flux factor can be obtained  by
 straightforward MD initialized in $A$ and
 counting the positive crossings through $\lambda_1=\lambda_0$ during the
 simulation run:
\begin{eqnarray}
\frac{\left \langle \phi_{1,0} \right \rangle}{ \left \langle
h_{\mathcal A} \right \rangle }= \frac{1}{\Delta
t}\frac{N_c^+}{N_{\textrm{MD}}}
\label{flxMD}
\end{eqnarray}
with $\Delta t$ the MD time step, $N_\textrm{MD}$ the number of MD steps, and
$N_c^+$ the number of counted positive crossings.  To
calculate the rate at constant temperature instead of constant energy,
one  can apply a 
Nos\'e-Hoover \cite{Nose84_1,Nose84_2,Hoov85,Martyna96} or Andersen
\cite{Andersen80} thermostat. However, one should be aware that these
thermostats do give the correct canonical distribution at a given
temperature, but modify the dynamics in an unphysical way.  Hence,
static averages $\left \langle A(x) \right \rangle$ will be correct,
but time correlation functions $\left \langle A(x_0) B(x_t) \right
\rangle$ most likely not.  As $N_c^+ \sim \left \langle
\theta\big(\lambda_1-\lambda(x_0) \big) \theta\big(\lambda(x_{\Delta
t})-\lambda_1 \big) \right \rangle$ is actually a correlation function
over a very short time, this effect will be small. However, if
necessary one can easily correct for this by explicitly counting
only phase points $x$ that in absence of the thermostat will cross $\lambda_1$ 
in one $\Delta t$ time-step. Applying this correction is
computationally cheap as it does not require any additional force
calculations.
In Appendix A we describe some possibilities for  further optimization of the flux algorithm.

\subsection{The path sampling algorithm}
\label{secshooting}
To calculate the conditional probabilities in TIS and PPTIS we use a
path sampling algorithm~\cite{Dellago02}.  However, there are some
differences with the classic TPS algorithm. Most importantly, in (PP)TIS the path 
length is variable, which has  a small implication for the acceptance criterion for the shooting move. In appendix B we derive this acceptance rule for arbitrary  (stochastic or deterministic) dynamics.
The main tools 
in the MC sampling of trajectory space are  the shooting move  and the time-reversal move~\cite{Dellago02}.  In particular for PPTIS  time-reversal moves can be quite effective. Shifting moves that enhanced statistics in TPS are not needed and even useless in (PP)TIS.

{\bf TIS algorithm:} The quantity of interest in TIS is the crossing
probability $P_A(\lambda_{i+1}|\lambda_i)$ (or 
$P_B(\lambda_{i-1}|\lambda_i)$ for the reverse rate constant $k_{BA}$).
To calculate this probability by sampling in the $\lambda_i$ interface ensemble one needs  an initial path that
starts in $A$ (at $\lambda_0$), crosses the interface $\lambda_i$ at
least once, and finally ends by either crossing $\lambda_0$ or
$\lambda_{i+1}$.  
In general one can take simply a successful path from the previous
$\lambda_{i-1}$ interface ensemble that reached $\lambda_{i}$,
and complete its evolution till reaching either $A$ or $\lambda_{i+1}$.
For more details on initial path generation we refer to
Ref.~\cite{Dellago02}).
The phase space point $x_0$ is then defined
as the first crossing point of this path with interface
$\lambda_i$.  It is convenient to use a discrete time index
$\tau=\textrm{int}(t/\Delta t)$, and let $\tau^b\equiv
\textrm{int}(t_{0}^b(x_0)/\Delta t)$ and $\tau^f\equiv
\textrm{int}(\textrm{min}[ t_0^f(x_0),t_{i+1}^f(x_0)]/\Delta t)$ be
the backward and forward terminal time slice indices, respectively.  Including
$x_0$, the initial path then consists of $N^{\rm (o)}=\tau^b+\tau^f+1$
time slices. Choosing a probability
$\gamma <1$ and a Gaussian width $\sigma_w$ we now start
following loop:

\begin{itemize}
\item Main loop
\begin{enumerate}
\item Take a uniform random number $\alpha_1$ in the interval $[0:1]$.
\item If $\alpha_1 < \gamma$ perform a time-reversal move. Otherwise,
perform a shooting move.
\item If the trial path
generated by either the time-reversal or shooting move is a proper path in the
$\lambda_i$ ensemble accept the move and replace the old path by the new one, 
otherwise keep the old path. Update averages and repeat from step 1.
\end{enumerate}
\item Time-reversal move
\begin{enumerate} 
\item If the current path ends at $\lambda_{i+1}$ reject the time-reversal move
and return to the main loop.
\item If the current  path starts and ends at $\lambda_0$, reverse the
momenta and the order of time-slices. 
On this reverse path, $x_0$ is
the new first crossing point with $\lambda_i$. Return to the main loop.
\end{enumerate}

\item Shooting move
\begin{enumerate}
\item On the current path with length $N^{\rm (o)}$ choose a random time
slice $\tau'$, with $ -\tau^b \leq \tau' \leq \tau^f$.
\item Change all  momenta of the particles at time-slice $\tau$ by adding 
small 
randomized displacements 
$\delta p=\delta w
\sqrt{m}$ with $\delta w$ taken from a Gaussian 
distribution with
width $\sigma_w$ and $m$ the mass of the particle~\cite{Dellago02}.
\item\label{alg1} In case of constant temperature (NVT) simulations:
accept the new momenta with a probability~\cite{FrenkelSmit}:
\begin{eqnarray}
\textrm{min} \Bigg[ 1,\exp\Big( \beta \big( E( x_{\tau' \Delta
t}^{\rm (o)}) -E(x_{\tau' \Delta t}^{\rm (n)}) \big) \Big) \Bigg]. \nonumber
\end{eqnarray}
Here, $E(x)$ is the total energy of the system at phase space point
$x$.  In case of constant energy (NVE) simulations in which possibly
also total linear- or angular momentum should be conserved: rescale
all the momenta of the system according to the procedure described in
Ref.~\cite{Geissler99} and accept the new rescaled momenta.

If the new momenta are accepted continue with step 4, else reject the
whole shooting move and return to the main loop.

\item Take a uniform random number $\alpha_2$ in the interval $[0:1]$
and determine a maximum allowed path length for the trial move by:
\begin{equation}
N_{\textrm{max}}^{\rm (n)}=\textrm{int}(N^{\rm (o)}/\alpha_2). \nonumber
\end{equation}

\item Integrate equations of motion backward in time by reversing the
momenta at time slice $\tau'$, until reaching either $\lambda_0$,
$\lambda_{i+1}$ or exceeding the maximum path length
$N_{\textrm{max}}^{\rm (n)}$.  If the backward trajectory did not reach
$\lambda_0$  reject and go back the main loop. Otherwise continue
with step 6.
\item Integrate from time slice $\tau'$ forward until reaching either
$\lambda_0$, $\lambda_{i+1}$ or exceeding the maximum path length
$N_{\textrm{max}}^{\rm (n)}$.  Reject and go back to the main loop if the
maximum path length is exceeded or if the entire trial path has no
crossing with interface $\lambda_i$. Otherwise continue with the next
step.

\item Accept the new path, reassign $x_0$ to be the first crossing
point with $\lambda_i$ and return to the main loop.
\end{enumerate}
\end{itemize}

Finally, the probability $P_A(\lambda_{i+1}|\lambda_i)$ follows from:
\begin{equation}
P_A(\lambda_{i+1}|\lambda_i)=\frac{N_p(0 \rightarrow
i+1)}{N_p(\textrm{total})}
\end{equation}
with $N_p(0 \rightarrow i+1)$ the number of sampled paths that connect 
$\lambda_0$ with $\lambda_{i+1}$ and $N_p(\textrm{total})$ the total
number sampled paths in the ensemble of interface $\lambda_{i}$.  

Time reversal moves do not require any force calculations. 
On the other hand two subsequent time reversals will just result in  the same 
path. Therefore, we usually take $\gamma=0.5$ giving
shooting and time reversal move an equal probability.  Similar
reasoning is applied to the choice of $\sigma_w$. If $\sigma_w$ is large,
many trial moves will
fail to create a proper path.
On the other hand a too small value of
$\sigma_w$ will result in almost the same path.  Practice has shown
that an optimal value of $\sigma_w$ is established when approximately
$40 \%$ of the paths is accepted~\cite{TPS99}. This will usually imply
that $\sigma_w$ will be  larger for the interfaces
$\lambda_i$ close to $A$ than the ones closer to $B$.  The mass
weighted momenta change at step 2 of the shooting algorithm is chosen
such that the velocity rescaling at step 3 maintains
detailed balance \cite{Geissler99}. In principle, NVT simulations
do not require rescaling and $\delta p$ can be taken from any
symmetric distribution. The integration of the equations of motion at
step 5 and 6 of the shooting move are normally performed by constant
energy MD simulations without using a thermostat to describe
the actual dynamics as realistic as possible. The temperature
only  appears at the acceptance criterion at step 3. 
In this algorithm we go from one
phase point $x_0^{\rm (o)}$ to a new one $x_0^{\rm (n)}$ by means of many MD
steps. Therefore, it has a strong similarity with hybrid MC
\cite{Duane87}.  Hence, the argument that the dynamics should be time
reversible and area preserving \cite{FrenkelSmit} should also be
applied here.  
For this reason, we strongly advice to use the velocity
Verlet \cite{Swope82} algorithm rather than higher order schemes such
as Runga-Kutta.  The maximum allowed path length
$N_{\textrm{max}}^{\rm (n)}$ in step 4 is introduced to maintain detailed
balance when sampling paths of different length and to avoid having to
reject  very long trial paths afterward \cite{ErpMoBol2003}.

{\bf PPTIS algorithm:}
the four one-interface probabilities
$p_i^\pm, p_i^=, p_i^\mp$, and $p_i^\ddagger$ for a single interface
$\lambda_i$ can be calculated simultaneously \cite{MoBolErp2004}
with paths that start at $\lambda_{i-1}$ or $\lambda_{i+1}$ and end by
crossing either $\lambda_{i-1}$ or $\lambda_{i+1}$. All paths
should have at least one crossing with $\lambda_i$.  Hence,
$\tau^b\equiv \textrm{int}(\textrm{min}[
t_{i-1}^b(x_0),t_{i+1}^b(x_0)]/\Delta t)$ and $\tau^f\equiv
\textrm{int}(\textrm{min}[ t_{i-1}^f(x_0),t_{i+1}^f(x_0)]/\Delta
t)$. The path sampling is then identical to the TIS algorithm except that 
$\lambda_{i-1}$ is used 
instead of $\lambda_0$,  time reversal moves are always
accepted and the backward integrating at step 5 is not rejected when
reaching $\lambda_{i+1}$ as paths  may start from both
sides.  The one-interface crossing probabilities are then given by
\begin{eqnarray}
p_i^{\pm}&=&\frac{N_p(i-1 \rightarrow i+1)}{ N_p(i-1 \rightarrow
i+1)+N_p(i-1 \rightarrow i-1)} \nonumber \\ p_i^{\mp} &=&\frac{N_p(i+1
\rightarrow i-1)}{ N_p(i+1 \rightarrow i-1)+N_p(i+1 \rightarrow i+1)}
\nonumber \\ p_i^= &=& 1-p_i^{\pm}, \qquad p_i^\ddagger = 1-p_i^{\mp}
\end{eqnarray}

\subsection{Defining the interfaces}
The order parameter  $\lambda$ in TPS and
TIS does not have to correspond to a reaction coordinate
that captures the essence of the reaction mechanism. The only
requirement is that $\lambda$ can distinguish between the two basins of 
attraction.  In TIS this occurs via the two outer interfaces
$\lambda_0$ and $\lambda_n$ that define state $A$ and $B$. The
definitions of $A$ and $B$ are more strict than in TPS
\cite{ErpMoBol2003}.  The boundaries $\lambda_0$ and $\lambda_n$
should be defined such that each trajectory between the stable states
is a rare event for the reaction we are interested in. In
addition, the probability that after this event the reverse reaction
occurs shortly thereafter must be as unlikely as an entirely new
event.  In other words, a trajectory that starts in $A$ and ends in
$B$ is allowed to leave region $B$ shortly thereafter, but the chance
that it re-enters region $A$ in a short time must be highly unlikely. 
Sometimes it is not sufficient for a proper definition of the boundaries 
$\lambda_0$ and $\lambda_n$ to only  use  configuration space.  In
the dimer study of Ref.~\cite{ErpMoBol2003} 
an additional kinetic energy 
constraint was introduced to ensure the stability of state $A$ and $B$.   

The intermediate interfaces can be chosen freely and should be placed to
optimize the efficiency. 
This is, of course, system dependent, but reasonable estimates
can be made a priori. 
Let us write down the total computation time as
$\textrm{CPU} \sim \sum_{i=1}^{N_W} N_i L_i$ with $N_W$ the number of windows
(interface ensembles), $N_i$ the number of paths in the ensemble of interface
$i$ required to obtain a desired precision 
$\epsilon_i$, and $L_i$ the average path length.  
Here, we 
neglect
the influence of rejections and the fact that two successive pathways
in the MC sequence are not completely uncorrelated. 
We chose the interface separations and the number of paths 
such that $P(\lambda_{i+1}|\lambda_i)=p$ and $N_i=n_p$,
resulting in $\epsilon_i=\epsilon$ for all $i$. 
The total error $\epsilon_{\rm tot}$, that we fix, 
is related by $\epsilon_{tot}^2=N_W \epsilon^2$ with
$\epsilon^2 \sim (1-p)/(p n_p)$.
Hence, $n_p \sim N_W (1-p)/p$ yielding 
$\textrm{CPU} \sim \sum_{i=1}^{N_W} L_i N_W (1-p)/p$.
The number of windows follows from 
$p^{N_W}= P(\lambda_{n}|\lambda_0) \Rightarrow
N_W \sim -1/\ln(p)$. Except for diffusive barrier 
crossings~\cite{MoBolErp2004},
that are most conveniently treated by PPTIS, the average 
path length $L_i$ has a linear dependence 
$\sim i (\lambda_n-\lambda_0)/N_W$~\cite{ErpMoBol2003}.
Taking this all into account, the final result gives 
$\textrm{CPU} \sim \ln(p)^{-2} (1-p)/p$ that has a minimum for
$p=0.2$.
Although, we made several assumptions in this derivation,
we believe that in general $P(\lambda_{i+1}|\lambda_{i})\approx 0.2$
for all $i$ is close to an optimum efficiency.

Between the interface positions one
can
use of a finer grid of sub-interfaces 
to obtain
the crossing probability
function $P_A(\lambda|\lambda_1)$ \cite{ErpMoBol2003} which is the path space 
analogy to a Landau free energy profile $F(\lambda)$. 
For PPTIS different requirements exist for the
position of interfaces. As the PPTIS formalism is based on a complete
memory loss over 
distances larger
than the interface separations, the
PPTIS interfaces should be set sufficiently far apart.  The
calculation of memory loss functions can help to determine the minimum
required distance to establish this \cite{MoBolErp2004}.

We would like to stress that although PPTIS transforms the system into a (pseudo) Markovian
hopping sequence based on local transition probabilities, it still maintains 
considerable history dependence. For example, the chance to go from interface $i$ to interface 
$i+1$ is assumed to be equal for 
the path that arrived at $i$ via the sequence $i-2 \rightarrow i-1 \rightarrow i$ or 
via the sequence $i \rightarrow i-1 \rightarrow i$. However, this transition
to $i+1$ from $i$ can still be different when its history 
had hopping sequence $i+1 \rightarrow i$.

\section{Improving the sampling}

\subsection{Parallel path swapping}
\label{parpathswap}
Biased sampling methods such as constrained dynamics~\cite{Ciccotti91}, 
multicanonical~\cite{Berg92} or umbrella sampling \cite{TV74,CCH89} can suffer 
from substantial ergodicity problems when the order parameters are not equal 
to the reaction coordinate. This lack of ergodicity usually shows up in 
hysteresis in the free energy curves (see e. g. ~\cite{ErpMeij04,Ensing2001}), 
and gives, besides a low transmission coefficient, rise to an additional error 
in the rate constant estimate.

Transition path sampling  was precisely devised to avoid this problem with
reaction coordinates, and, in a way, also avoids ergodicity problems 
due to the non-local nature of the shooting move. This advantage showed up  in the water trimer study \cite{Geissler99} where the TPS algorithm 
was capable of finding two reaction mechanisms across different saddle points 
separated by a barrier higher than the total energy of the NVE simulation.
We stress that this would have been much more difficult to achieve or even impossible in an umbrella 
sampling algorithm with several narrow windows. 
However, path sampling can also suffer from ergodicity 
problems if large barriers separate multiple reaction channels in a high 
dimensional rough energy landscape. 
In particular in the case of PPTIS, the short paths are much less likely to 
overcome such barriers.

Parallel tempering techniques (also known as Replica Exchange methods)
can facilitate the sampling \cite{Vlugt2001}, but
requires a rather large computational effort and cannot be applied at 
constant energy.  Here, we propose a less expensive parallel method  especially tailored for PPTIS to enhance 
ergodicity.  This  parallel path swapping 
(PPS) technique is based on the exchange of paths between two subsequent 
interface ensembles.
\begin{figure}[t]
\begin{center}
\includegraphics[width=10cm,keepaspectratio]{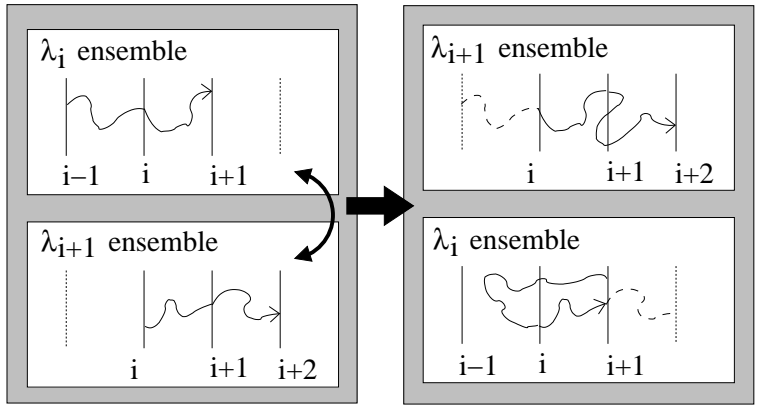}
\end{center}
\caption{Path swapping move for PPTIS. The last half of the path in the 
$\lambda_i$ ensemble
and the first half of the path in the $\lambda_{i+1}$ are swapped 
to the $\lambda_{i+1}$ and $\lambda_{i}$ ensembles, respectively.}
\label{figpathswap}
\end{figure}
Fig.~\ref{figpathswap} shows one path in 
the $\lambda_i$ ensemble, consisting of all possible paths crossing $\lambda_i$ while  starting and ending at either
$\lambda_{i- 1}$ or $\lambda_{i+1}$, 
and one in the $\lambda_{i+1}$ ensemble consisting of all paths 
crossing $\lambda_{i+1}$ at least once, while starting and ending at  either 
$\lambda_i$ or $\lambda_{i+2}$.
We introduce a new MC move that attempts  swapping the current path of 
the  $\lambda_i$  ensemble with that  of the $\lambda_{i+1}$-ensemble, as depicted in
Fig.~\ref{figpathswap}.  
The swap move will  be rejected if the $\lambda_i$
ensemble path does not end at $\lambda_{i+1}$ or if the  
$\lambda_{i+1}$ ensemble path does not start at $\lambda_i$.
Otherwise, the move is accepted and the two  trajectories are swapped from one 
ensemble to the other. 
Integrating the equations of motion backward (for the $\lambda_i$ ensemble)
and forward (for the $\lambda_{i+1}$ ensemble) will result in two entirely 
new paths for both ensembles. 
The acceptance/rejection criterion
appears before any expensive computation of MD trajectories.
Moreover, once accepted we  obtain  
a new path for both ensembles for price of effectively only one path. 
This makes the path swapping move useful even if for  systems
 not suffering from ergodicity problems.

Another advantage of PPS is that it allows to go beyond the pseudo-Markovian 
description of PPTIS. Fig.~\ref{figpathswap} shows that the paths at the right hand side, 
if we include the dashed trajectory part, can connect four interfaces instead of only  three. 
This extension  allows for a long range verification of the memory loss assumption. 
Also, the development of new, smart algorithms based on PPS might be able to correct for memory effects or  to  search  for ideal interface positions on the fly.  

While PPS is very effective when the confinement of short paths in PPTIS can cause sampling problems,
even TIS and TPS algorithms   might benefit from path swapping
when multiple reaction channels exist. 

\subsection{CBMC based shooting moves}
Originally developed to sample polymers at high densities, the
Configurational Bias Monte Carlo (CBMC) technique grows chain molecules in a
biased fashion in order to avoid unfavorable overlap of the
beads~\cite{Rosenbluth55,Siepmann92,dePablo92,Frenkel92}.  The
similarity between growing polymers and generating dynamical
trajectories was the inspiration for the development of TPS 
and has been exploited in the sampling of the stochastic
path action~\cite{TPS98,Felix98}. However,
this CBMC-like technique was found to be less
effective than the shooting algorithm~\cite{Dellago02}.  Here, we
propose a combination of the shooting move with CBMC for diffusive
systems that suffer from low acceptance due to a non flat rough free
energy barrier.  When shooting from one basin of attraction in such
systems, the Lyapunov instability causes the paths to diverge and
return to the same basin of attraction before crossing the barrier. The
use of some stochastic noise allows shooting in only one time
direction and alleviates this problem slightly~\cite{BolhuisJPCM,BolhuisPNAS}, 
but at the price that independent
pathways are generated only after a number of accepted shooting moves
from the barrier region. This slow exploration of path space is even
worse for processes proceeding via multiple dynamical bottlenecks, for
instance reactions taking place though a short lived intermediate
state 

\begin{figure}[t]
\begin{center}
\includegraphics[width=7.5cm,keepaspectratio]{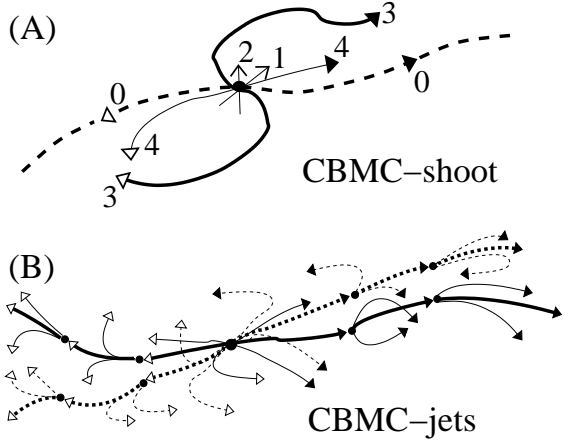}
\end{center}
\caption{CBMC shooting move. (A) At the shooting point
at the old trajectory (dashed line) 
four trial segments are released. In this example
the momenta of
segment 1 and 2 have been rejected and are not integrated further.
Segment 0 is retracing the old path.  Of the trial segments, segment 
3 has come farthest  in its forward (solid arrow) 
and backward (open arrow) time evolution and will consequently have the highest
weight. (B) The use of stochasticity allows the creation of trajectory jets at several points along the path.  
At each junction the path will follow the 
most favorable direction (bold solid line). 
The creation of trajectory jets at the old path (bold dashed line)
is required to maintain super-detailed balance.}
\label{figCBMC}
\end{figure}

Within the shooting algorithm, CBMC can be applied both at the shooting
point (the random time slice for which we change the momenta) and
along the path by introducing some stochastic noise.  At the shooting
point  $\tau'$  we generate a set of $N_s$ momenta displacements 
$\{ \delta p ^{\rm (n)}\}$, and accept these displacements using step 3 in Sec.~\ref{secshooting}.    
Each phase point is then  integrated 
forward and backward for a time $\tau_L$, resulting in $N_s$ trajectory 
segments ${\bf s}_j \equiv \{x^{(j)}_{(\tau' -\tau_L) \Delta t}, \ldots, 
x^{(j)}_{(\tau' +\tau_L) \Delta t} \}$, for  $j=1, \ldots, N_s$
(See Fig.~\ref{figCBMC}-A). The time
interval $\tau_L$ should be large enough to decide whether a
trajectory has a chance of being successful, but much smaller than the
average path length of a complete trajectory. All  path segments are
given a weight $w_j$ 
\begin{eqnarray}
w_j^{\rm (n)}=\Psi\big(\delta p^{\rm (n)} \big) {\mathcal F}({\bf s}_j)
\label{CBMCweight}
\end{eqnarray}
where $\Psi$ equals 1 (else 0) for accepted momenta changes $\delta p$
at the shooting point $\tau'$.
The biasing function ${\mathcal F}$
should be chosen to give the highest weight $w_j$ to those segments
that are most likely to produce a complete path of the corresponding
interface ensemble.  One possibility is to choose 
${\mathcal F} 
=\exp(\alpha \Delta \lambda)$ with $\Delta
\lambda=\lambda(x_{(\tau'+\tau_L)\Delta t})-\lambda(x_{(\tau'-\tau_L) \Delta t})$
and $\alpha$ a parameter optimized to the steepness of the barrier at
$x_{\tau' \Delta t}$. In that case, ${\mathcal F}$   is a function only
of the backward and forward end points of the path segments ${\bf s}_j$. 
The Rosenbluth factor for the set of trajectory segments is
\begin{eqnarray}
W^{\rm (n)} \equiv \sum_{j=1}^{N_s} w_j
\end{eqnarray} 
One of the segments ${\bf s}_i$ is selected with a probability
$w_i/W^{\rm (n)}$.  To correct for this bias and to obey detailed balance, we also have to calculate the Rosenbluth factor $W^{\rm (o)}$ for the old
path. The procedure is the same
as above, but now we apply
$N_s-1$ new random momenta changes
$\{\delta p^{\rm (o)} \}$ to the momenta of ${\bf s}_i$ at the 
same shooting point 
and again generate a set of segments of length $2 \tau_L \Delta t$.
This set is completed by adding
segment ${\bf s}_0$ of the same length from the old path.
The  Rosenbluth factor for the old path equals
\begin{eqnarray}
W^{\rm (o)} \equiv \sum_{j=0}^{N_s-1} w_j^{\rm (o)}.
\end{eqnarray}
where
$w_0^{\rm (o)}$ is the weight of segment ${\bf s}_0$ , and $w_j^{\rm (o)}$ with
$j=1,\ldots, N_s-1$ are the weights for the segments  that follow
from $\{ \delta {p}^{\rm (o)} \}$.

By imposing super detailed balance \cite{FrenkelSmit} the
acceptance probability of segment $i$ becomes
\begin{eqnarray}
P_\textrm{acc}(\textrm{\bf s}_0 \rightarrow \textrm{\bf s}_i)
=\min\left[1,\frac{w_0^{\rm (o)} W^{\rm (n)} \rho(x_{\tau' \Delta t}^{\rm (n)}) }{w_i^{\rm (n)} W^{\rm (o)} \rho(x_{\tau' \Delta t}^{\rm (o)})}\right].
\label{CBMCshoot}
\end{eqnarray}
Here, the weight functions $w_i$  and the distributions $\rho$
are still present, because they do not cancel as in the  standard CBMC expression.  The accepted segment is
integrated to the complete path just as in the normal shooting move of
Sec.~\ref{secshooting}.
Of course, this procedure is computationally more expensive than the
standard shooting move. However, the biasing function ${\mathcal F}$
allows to choose a segment with much higher probability to become a
accepted path. We expect an increase in sampling efficiency when
the gain in acceptance outweighs the cost of the construction of the
trajectory segment sets.

In the above algorithm we only can bias the growth of the first
segment of the trajectory (the analog of the polymer in standard CBMC)
because the rest of the trajectory follows deterministically once the
first segment has been chosen. In the standard polymer CBMC a bias is
introduced at each segment, and we can make use of the full power of 
CBMC if we consider stochastic trajectories. Introducing a small
amount of stochasticity by for instance the Andersen thermostat
\cite{BolhuisJPCM} or by making use of the periodic boundary
condition\cite{Ciccotti80} will hardly change the dynamical properties of the transition process.

Stochasticity allows us to create trajectory jets at several
points along the paths (See Fig.~\ref{figCBMC}-B).
The first segment is created as in the deterministic procedure above.
However, the chosen segment is not integrated to the full path
length. Instead, we start with the end point of the forward trajectory
and integrate a 'jet' of forward trajectory segments each evolving
differently according to its own random noise.  Each segment $j$ of
this 'jet' $k$ has a weight $w_{jk}$ similar to Eq.~(\ref{CBMCweight})
and each jet will have a total weight $W_k=\sum_j w_{jk}$.  We select
a segment $i$ according to its relative weight $w_{ik}/W_k$, and
continue with the next jet of forward segments.  The same is done for
the backward paths, until the path is completed.  After generating the
new path, we have to repeat the 'jet' procedure for the old path as
depicted in Fig.~\ref{figCBMC} in order to calculate the
Rosenbluth factor of the old path.
The total Rosenbluth factors are now
\begin{eqnarray}
W_{RF}^{\rm (n)} = \prod_k W_k^{\rm (n)}, \qquad W_{RF}^{\rm (o)}= \prod_k W_k^{\rm (o)}
\end{eqnarray}
where $k$ runs over all the jets including the one at the shooting
point $x_{\tau' \Delta t}$. The final acceptance criterion obeying super detailed
balance is then
\begin{eqnarray}
P_\textrm{acc}(o\rightarrow n)=
\textrm{min}\left[1,\frac{ W_{RF}^{\rm (n)} \rho(x_{\tau' \Delta t}^{\rm (n)}) \prod_k w_{0k}^{(o)}}{ W_{RF}^{\rm (o)} \rho(x_{\tau' \Delta t}^{\rm (o)}) \prod_k w_{ik}^{\rm (n)}}
\right]
\end{eqnarray} 
where $w_{0k}^{\rm (o)}$ is the segment weight at jet $k$  on the old
path and $w_{ik}^{\rm (n)}$ is the weight of the selected segment of jet $k$ on the new path.
To take into account the change in path-length one should include a factor
$\textrm{min}[1,N^{\rm (o)}/N^{\rm (n)}]$, but this is usually implemented by
defining a maximum path length as explained at step 4 of the shooting
algorithm in Sec.~\ref{secshooting}. The above algorithm could be
useful when the standard shooting move suffers from extreme low
acceptance ratios.

\subsection{Time as transition parameter}
In TIS the choice of the  order parameter is not critical
as $\lambda$  does not have to correspond to the  
reaction coordinate. Yet, it is possible that 
the order parameter $\lambda$ can  bias the outcome of transition mechanism 
and rate constants 
-- although much less than  for the TST reactive flux method--, 
for instance, when the reaction mechanism leads in a direction that 
$\lambda$ does not allow.
In principle, an order parameter-free 
sampling method is, therefore, highly desirable when examining unexpected
contra-intuitive reaction mechanisms.  
One possibility for such a bias-free method is by using the 
time on the path outside $A$ as transition parameter 
(we use 'transition' instead of 'order' to indicate that it is not a 
traditional order parameter as it is not based on a phase point).

For a  particular stable state $A$ definition $\lambda_0$, 
$P_A({\mathcal T}_{i+1}|{\mathcal T}_{i})$ is the probability that a path, starting
from  $\lambda_0$ and remaining outside $A$ over a time ${\mathcal T}_i$,
remains even longer outside $A$ until at least ${\mathcal T}_{i+1} > {\mathcal T}_i$.
To calculate the probability $P_A({\mathcal T}_{i+1}|{\mathcal T}_{i})$  by a bias-free 
TIS simulation we generate an ensemble of 
trajectories that have path lengths between ${\mathcal T}_{i}$ and ${\mathcal T}_{i+1}$ using the
shooting algorithm of Sec.~\ref{secshooting}. 
At the shooting point, we integrate backward until reaching $\lambda_0$ or
until the length of the trial trajectory exceeds ${\mathcal T}_{i+1}$ 
(or $N_{\rm max}^{\rm (n)} \Delta t$ as defined at step 4 of 
the shooting algorithm).
If the backward trajectory exceeds either ${\mathcal T}_{i+1}$ or 
$N_{\rm max}^{\rm (n)} \Delta t$ the shooting move is rejected.
The forward trajectory is continued until
reaching $\lambda_0$, or until a path length of ${\mathcal T}_{i+1},$ or $N_{\rm max}^{\rm (n)} \Delta t$. The trial path is rejected 
if $N_{\rm max}^{\rm (n)} \Delta t$ 
is exceeded or if the trajectory ends at $\lambda_0$
in a time shorter than ${\mathcal T}_i$. 
In the subsequent ensemble, the probability $P_A({\mathcal T}_{i+2}|{\mathcal T}_{i+1})$ 
for ${\mathcal T}_{i+2}>{\mathcal T}_{i+1}$ is calculated for all paths with at least a length 
${\mathcal T}_{i+1}$.

\begin{figure}[ht!]
\begin{center}
\includegraphics[width=10cm,keepaspectratio]{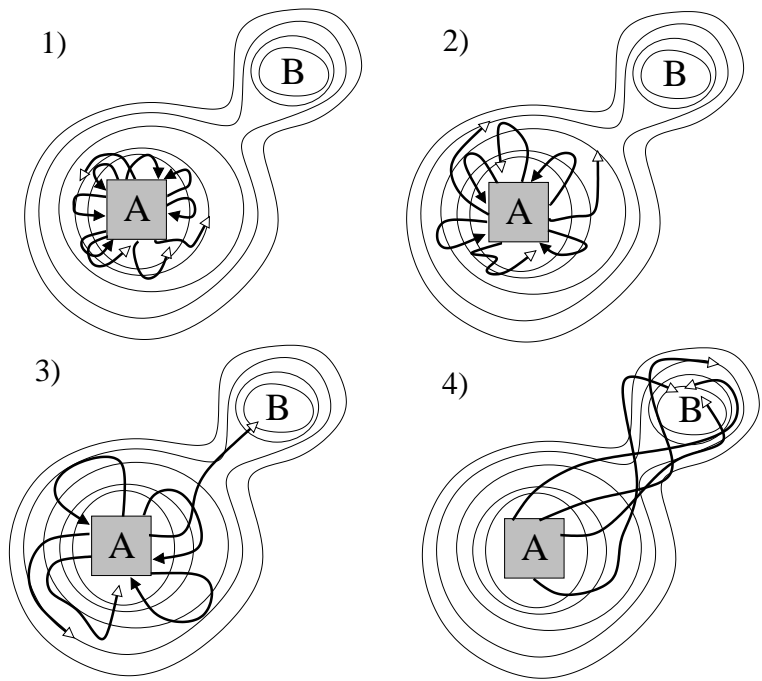}
\end{center}
\caption{Time as transition parameter.
The square denotes the definition of the boundary for state $A$.
The thin lines are free energy contour lines.
The four panels show the representation of generated trajectories
in successive time-interface ensembles. At panel 1), $P_A({\mathcal T}_{i+1}|{\mathcal T}_{i})$
is the fraction of of trajectories that stay outside $A$ longer than
${\mathcal T}_{i+1}$ (open arrows). All trajectories have at least a length
${\mathcal T}_i$. The solid arrows  are the paths that return to $A$ before ${\mathcal T}_{i+1}$.
At panel 2), $P_A({\mathcal T}_{i+2}|{\mathcal T}_{i+1})$ is calculated for paths that remain outside
$A$ longer than ${\mathcal T}_{i+1}$. The minimum length of the paths
is further increased at panel 3). Incidentally, a path will end up in the
yet unknown state $B$. At panel 4) the minimum path length
constraint forces all the paths into the metastable state region $B$.
From here, they will not return. Hence $P_A({\mathcal T}|0)$ will show a plateau.}
\label{figtimepar}
\end{figure}

This method, as illustrated in 
Fig.~\ref{figtimepar}, 
will thus explore automatically the regions further
and further outside $A$. At some moment it will find the closest
stable state region (state $B$). Trajectories reaching this region 
will not go back to $A$, hence, the overall 
crossing probability function $P_A({\mathcal T}|0)$ will show a plateau at some time ${\mathcal T}$
similar to standard TIS.

Two-ended path sampling methods, such as TIS, PPTIS  and TPS can  only treat 
processes in which both stable states $A$ and $B$ are known. They cannot find the final state starting from a single stable state, a fact already 
discussed by Dellago and Chandler \cite{DellagoSIMU}.
The algorithm  described here, might be a  solution to this problem.

\section{Extracting information from path ensembles}

\subsection{Reaction mechanism}
The ensemble of paths collected by the TIS algorithm can be used to
investigate the reaction mechanism. 
We believe that for this purpose the TIS path ensembles might even be more 
useful than the TPS path ensembles. 
The TPS method, first samples paths that all successfully reach $B$ in the part to obtain the reactive flux
 function and then in the second step samples artificially short trajectories of fixed length to calculate the time correlation function C(t). Because of this constraint, the resulting ensembles do not give useful information about the reaction. 
The TIS $\lambda_i$-ensembles, on the other hand, contain the correct distribution
of paths that have crossed $\lambda_i$ and are either going on to $\lambda_{i+1}$ or return to A.
Some hidden order parameters can only be discovered
by carefully comparing configurations along  reactive and unreactive trajectories
that are similar in terms of order parameters which at first sight were
considered as  being the (only) important ones.  
For instance, the comparison of reactive and unreactive geometries with an almost identical
orientation of the reactants showed that
precise tetrahedral ordering of the solvent water molecules was an important factor 
in the hydration reaction of ketones \cite{ErpMeij04}. 
Although  there is currently no
systematic way to extract the reaction coordinates from a path ensemble, once
a reaction coordinate is postulated based on physical insight it can be tested using  
committor distributions~\cite{Bolhuis02}.

\subsection{Activation energies}
The  activation energy $E_a$ is an important experimentally accessible quantity and is defined by the Arrhenius law
\begin{eqnarray}
k = A e^{-\beta E_a},
\label{Arrhenius}
\end{eqnarray}
where A is a system dependent prefactor. 
In fact, $A$ and $E_a$ may also be temperature 
dependent. Such non Arrhenius behavior can be quite severe: 
sometimes  reaction rates are even decreasing with 
increasing temperature, resulting in a 'negative activation energy'
(see e.g. \cite{Shimomur67}).
>From Eq.~(\ref{Arrhenius}) it follows  that
\begin{eqnarray}
\label{Eader}
E_a=-\frac{\partial \ln k_{AB}(\beta) }{\partial \beta}.
\end{eqnarray}
An algorithm to calculate $E_a$ in a TPS simulation was given in
Ref.~\cite{DellagoMOLSIM}. 
Here, we use a similar approach to calculate $E_a$ in a 
canonical TIS simulation. Substitution of 
Eq.~(\ref{kabphiP}) in Eq.~(\ref{Eader}) results in
\begin{eqnarray}
E_a &=& -\frac{\partial}{\partial \beta}
\Big[ \ln \left \langle \phi_{1,0} \right \rangle -  \ln
\left \langle h_{\mathcal A} \right \rangle 
+ \sum_{i=1}^{n-1} \Big(
\ln \left \langle \Phi_{i,0}^{i+1,0} \right \rangle - \ln
\left \langle \phi_{i,0} \right \rangle \Big) \Big]
\label{actEN1}
\end{eqnarray}
For any function $A(x)$ we can write
\begin{eqnarray}
-\frac{\partial \ln \left \langle A(x) \right \rangle }{\partial \beta}=
 \left \langle E(x) \right \rangle_A -  \left \langle E(x) \right \rangle
\end{eqnarray}
with 
$\left \langle E(x) \right \rangle_A=
\left \langle A(x) E(x) \right \rangle/\left \langle A(x) \right \rangle$.
Using 
\begin{eqnarray}
\left \langle  E(x) \right \rangle_{\Phi_{i,0}^{i+1,0}}=
\left \langle  E(x) \right \rangle_{\phi_{i+1,0}},
\end{eqnarray}
most terms in Eq.~(\ref{actEN1}) cancel, only leaving
\begin{eqnarray}
E_a =
\left \langle  E(x) \right \rangle_{\Phi_{n-1,0}^{n,0}}-
\left \langle  E(x) \right \rangle_{h_{\mathcal A}},
\end{eqnarray}
which  is the difference between the average energy of state $A$ and the 
energy of the transition pathways connecting $A$ with $B$.
Consequently, the calculation of the $E_a$ does not require
all  interface ensembles, but only the last  ensemble $\lambda_{n-1}$.
However, if all the path ensembles $i=1,\ldots, n-1$ are available
an activation energy function
\begin{eqnarray}
E_a(\lambda_i) =
\left \langle  E(x) \right \rangle_{\Phi_{i-1,0}^{i,0}}-
\left \langle  E(x) \right \rangle_{h_{\mathcal A}}
\end{eqnarray}
can be calculated that should converge to a plateau analogous to   
the crossing 
probability $P(\lambda|\lambda_1)$. 
A finer grid of sub-interfaces can be applied
to obtain a continuous smooth function $E_a(\lambda)$.

Again, there is a subtle difference between the TPS and TIS algorithms.
For the reaction rate determination, TPS requires a plateau  in the time 
correlation function of Eq.~(\ref{corrTPS}), while TIS should
give a plateau in $\lambda$ for the crossing probability 
$P(\lambda|\lambda_1)$. Similarly, the TPS activation energy is
expressed as a time dependent function that will converge to a plateau
at times $t={\mathcal T}$ 
\cite{DellagoMOLSIM}, while the TIS activation energy reaches a
plateau in terms of $\lambda$. 

Not all reactions show Arrhenius behavior. Therefore, it would
be interesting to determine  $k_{AB}(\beta)$ for a range of temperatures.
One can estimate the rate for a slightly different temperature 
by  reweighting the crossing probabilities\cite{TV74,CCH89}.
If $\left \langle A(x) \right \rangle$ is the average
of an observable $A(x)$ at inverse temperature $\beta$, then
$\left \langle A(x) e^{-\Delta \beta E(x)} \right \rangle/
\left \langle e^{-\Delta \beta E(x)}  \right \rangle$ should
be the average at inverse temperature $\beta+\Delta \beta$.
This reweighting technique can also be applied to the crossing 
probabilities~(\ref{P4int}) and the flux~(\ref{flxMD}).
Of course, $\Delta \beta$ should be small to obtain good enough statistics.
The calculation of the temperature dependence of  individual 
crossing probabilities has the advantage that the origin 
of possible non-Arrhenius behavior might be located (in terms of $\lambda$) 
along the reaction path.

\section{Summary and Conclusions}

We reviewed the basic concepts of TIS and PPTIS and explained their relation
to TST based methods and TPS. We believe that path sampling methods, TPS, TIS and PPTIS, are powerful 
when dealing with high dimensional complex process for which is a reaction coordinate is lacking.
Among these methods, TIS can be considered as an improvement upon the original TPS giving a complete
non-Markovian description of the reactive event, but more efficient. 
PPTIS improves the efficiency even more, but relies on the assumption of memory loss between 
interfaces. Hence, it should only be applied for diffusive barrier crossings.
In addition to this review,  we have introduced several new techniques in this paper.
These novel methods comprise the CBMC based shooting moves, order parameter free methods,
parallel path swapping and the calculation of activation energies.
The efficiency of these methods should be tested by future
simulations. We plan study this in the near future.

\ack 
We thank Daniele Moroni for useful discussions and carefully reading
this article.
T.S.v.E acknowledges
the support by a Marie Curie Intra-European Fellowships 
(MEIF-CT-2003-501976) within 
the 6th European Community Framework Programme and the support through
the European Network LOCNET HPRN-CT-1999-00163.  P.G.B
acknowledges support from the FOM (Stichting Fundamenteel Onderzoek
der Materie).

\appendix

\section{The flux revisited}
In some cases, we can improve the efficiency of the flux 
calculation by separating the flux into 
the probability to be on the  $\lambda_1$ surface  times a factor  integrating
over all possible velocities when leaving the surface.
The flux term can then be calculated by combining straightforward MD with,  as soon as we cross the $\lambda_1$ surface,
the sampling of  sets of randomized Gaussian distributed 
velocities $\dot{\lambda}$, after which 
 the MD trajectory is continued with the old original momenta.
In this way, we make optimal use of the statistics of the crossing points.
The velocity sampling does not require force calculations and is therefore cheap.
In the following we assume that we always take $\lambda_1=\lambda_0$.
Similar to Eq.~(\ref{rateTST}) we can write: 
\begin{eqnarray}
\label{eq:c1}
 \frac{\left \langle \phi_{1,0} \right \rangle }{
\left \langle h_{\mathcal A}  \right \rangle }=
\left \langle \theta(\dot{\lambda)} \dot{\lambda} \right \rangle_{\lambda_1}
P(\lambda_1)_{x \in {\mathcal A} }
\label{flx2}
\end{eqnarray}
with
\begin{eqnarray}
P(\lambda_1)_{x \in {\mathcal A} } \equiv
\frac{\left \langle  \delta(\lambda(x)-\lambda_1) \right \rangle}{
\left \langle h_{{\mathcal A} } \right \rangle}
\end{eqnarray} 
The two terms 
$\frac{\left \langle \phi_{1,0} \right \rangle }{
\left \langle h_{\mathcal A}  \right \rangle }$ 
and $P(\lambda_1)_{x \in {\mathcal A} }$ can be obtained in the same MD
simulation. As
$\left \langle \delta(\lambda(x)-\lambda_1) \right \rangle d\lambda $
is equal to the probability to find the system in the interval
$[\lambda_1 - \frac{1}{2} d \lambda : \lambda_1 + \frac{1}{2} d \lambda ]$,
it can be  measured 
by defining a width $d \lambda$ and performing 
a MD (or MC) simulation starting in $A$:
\begin{eqnarray}
\label{eq:pl}
P(\lambda_1)_{x \in {\mathcal A} }=
\frac{1}{d \lambda} \frac{N_{\lambda_1}}{N_\textrm{MD}}
\end{eqnarray}
with $N_{\lambda_1}$ the number of counts in the specified interval
and $N_\textrm{MD}$ the number of MD steps.
However, this number can depend sensitively on the choice of bin width $d \lambda$.
Ideally one would like  $d \lambda$ to be as small as possible, at the cost of having  
to perform a very long simulation run for a statistically accurate
number $N_{\lambda_1}$.
A better option is to weigh the crossings with a function
depending on the velocity. Assume that we cross $\lambda_1$ in one MD step from  
$x_{i \Delta t}$ to $x_{(i+1) \Delta t}$.  If $d \lambda$ is small 
neither of these points will lie inside the interval.
However, assuming a linear dynamics between these points, 
the system traverses from $x_{i \Delta t}$ to $x_{(i+1) \Delta t}$ in $N_\textrm{sub}$ equidistant
sub steps. The number of  phase points $N_{\lambda_1}$ that lie in the $d \lambda$ 
interval of this short linear trajectory is approximately  $d \lambda N_\textrm{sub}/ 
| \lambda(x_{(i+1) \Delta t})-\lambda(x_{i\Delta t}) |$. The total number
of MD moves $N_\textrm{MD}$, of course, also increases  by a factor $N_\textrm{sub}$.
So Eq.(~\ref{eq:pl}) becomes
\begin{eqnarray}
P(\lambda_1)_{x \in {\mathcal A} } =
\frac{1}{N_\textrm{MD}} 
{\sum_{i}}^* \frac{1}{|\lambda(x_{(i+1) \Delta t})-\lambda(x_{i \Delta t}) |}
\end{eqnarray}
where the $*$ indicates that the summation has to be performed only for
points $i$ along the trajectory for which $x_{i\Delta t} \rightarrow x_{(i+1) \Delta t}$ 
showed a crossing (positive or negative) with interface $\lambda_1$. Further optimization can 
be achieved by writing
 $|\lambda(x_{(i+1) \Delta t})-\lambda(x_{i\Delta t}) |= |\dot{\lambda}_{x_{i\Delta t}}  | \Delta t 
+ {\mathcal O}( \Delta t^2)= |\dot{\lambda}_{x_{(i+1)\Delta t}}  | \Delta t
+ {\mathcal O}( \Delta t^2)$, but the
velocity $\dot{\lambda}$ at the interface would give the most exact result.
If we also assume a linear change in time for the velocities between $i$ 
and $i+1$, our best estimate for 
$P(\lambda_1)_{x \in {\mathcal A} }$ is:
\begin{eqnarray}
P(\lambda_1)_{x \in {\mathcal A} } &=&
\frac{1}{N_\textrm{MD} \Delta t} {\sum_i}^* \frac{1}{
| \dot{\lambda}(x_{i \Delta t};\lambda_1) | }
\label{prob_est}
\end{eqnarray}
where we have introduced the notation $g(x_{i \Delta t};\lambda_j)$ to 
denote the function $g(x)$ at the crossing point of interface $\lambda_j$ obtained by  a linear interpolation of the function 
between two successive trajectory points 
$x_{i \Delta t} \rightarrow x_{(i+1)\Delta t}$:
\begin{eqnarray}
g(x_{i \Delta t};\lambda_j) & \equiv &
\frac{1}{\lambda(x_{(i+1)\Delta t})-\lambda(x_{i\Delta t}) }
\big\{  [\lambda(x_{(i+1)\Delta t})-\lambda_j ] 
 g(x_{i\Delta t})
+ [\lambda_j-\lambda(x_{i\Delta t})] g(x_{(i+1)\Delta t}) \big\} \nonumber \\
\end{eqnarray}
The factor 
$\left \langle \theta(\dot{\lambda}) \dot{\lambda} \right \rangle_{ \lambda_1}$
in Eq.~(\ref{eq:c1}) can be calculated in the same MD simulation with an additional sampling
procedure.
In some simple cases, there is even an analytically
expression.
For instance, in case the $x$-coordinate of particle $j$ is the order parameter,
$\lambda(x)=r_{jx}$ in a constant temperature (NVT) simulation,
we would obtain
$\left \langle \theta(\dot{\lambda}) \dot{\lambda} \right \rangle_{ \lambda_1}
=\frac{1}{\sqrt{2 \pi \beta m_j}}$ with $m_j$ the mass of this particle.
However, for  more complex $\lambda(x)$, such as the 
distance between two particles $i$ and $j$, 
$\lambda(x)=|r_i-r_j|$, no simple analytic expression exists.
The calculation of
$\left \langle \theta(\dot{\lambda}) \dot{\lambda} \right \rangle_{ \lambda_1}$
can then be  calculated by sampling a random set of $N_\textrm{MC}$ velocities
$\dot{\lambda}$ as soon  as a crossing is detected:
\begin{eqnarray}
\left \langle \theta(\dot{\lambda}) \dot{\lambda} \right \rangle_{ \lambda_1}
= \frac{{\sum_i}^* \Big[ \frac{1}{|\dot{\lambda}(x_{i \Delta t};\lambda_1)|}
\sum_j^{N_\textrm{MC}} \theta(\dot{\tilde{\lambda}}) \dot{\tilde{\lambda}} 
\Big] }{ N_\textrm{MC} {\sum_i}^*  
\Big[ \frac{1}{|\dot{\lambda}(x_{i \Delta t};\lambda_1)|}
 \Big]}
\label{sampleMDMC}
\end{eqnarray}
where $i$ runs over all MD crossings with interface $\lambda_1$,
$\dot{\lambda}(x_{i \Delta t};\lambda_1)$ is the MD crossing velocity
through $\lambda_1$, $j$ runs over the $N_\textrm{MC}$ 'artificial' velocities $\dot{\tilde{\lambda}}$ that are taken from a proper distribution $P(\dot{\tilde{\lambda}}|x_{i \Delta t})$.
For NVT simulations without additional
constraints this distribution $P(\dot{\tilde{\lambda}}|x_{i \Delta t})$ does not  depend
on the phase point $x_{i \Delta t}$ and we can simply sample $\dot{\tilde{\lambda}}(\{p\})$
where the  momenta $\{p\}$ defining $\dot{\lambda}$ are taken from a Gaussian distribution
For NVE simulations, the  distribution $P(\dot{\tilde{\lambda}}|x_{i \Delta t})$ does depend  $x_{i \Delta t}$ and 
we have to change all momenta and distribute them
on the hypersphere defined by the kinetic energy $K=E-V(x_{i \Delta t};\lambda_1)$
with $E$ the total energy and $V(x_{i \Delta t};\lambda_1)$ the total
potential energy at the crossing point.
The  proper sampling of momenta distributions in the presence of linear constraints, such
as linear and angular momentum  is explained in 
Ref.~\cite{Geissler99}. Clearly, if $P(\dot{\tilde{\lambda}}|x_{i \Delta t})
=\delta\big(\dot{\tilde{\lambda}}-\dot{\lambda}(x_{i \Delta t};\lambda_1) \big)$,
Eq.~(\ref{sampleMDMC}) would be equal to
$N_c^+/{\sum_i}^*|\dot{\lambda}(x_{i \Delta t};\lambda_1)|^{-1}$
leaving Eq.~(\ref{flx2}) identical to Eq.~(\ref{flxMD}) from which
we started.

\section{TIS  shooting  acceptance criterion for stochastic dynamics}
\label{stochdyn}
Although we assume throughout the paper that the equations of motion were     
deterministic, it  is sometimes useful to 
implement some stochasticity into the dynamics, or consider completely stochastic equation of  
motion  such as Brownian Dynamics \cite{Dellago02,BolhuisJPCM,BolhuisPNAS}.  
Quantities like $\bar{h}_{\mathcal A}(x_0)$ are, then, no longer 
just 1 or 0 but turn into probabilities with a fractional value.  
Moreover, for stochastic dynamics it is not trivial whether we are 
allowed to use the path 
that generated $x_0^{\rm (o)}$ as our instrument to search for the new 
phase point $x_0^{\rm (n)}$. In this appendix we derive the acceptance 
probability for the shooting algorithm for arbitrary dynamics along the same 
lines as in Ref.~\cite{Dellago02}. At start, we try to be as general 
as possible making the least possible assumptions on the type of dynamics
or on whether the system is in equilibrium or not.
For this purpose, it is most convenient to 
use  the path space description, instead of phase space. The weight or 
probability  density ${\mathcal P}[{\bf x}] $ for a single path ${\bf x}  \equiv
\{x_{ -\tau^b \Delta t },\ldots;x_0~;\ldots, x_{ +\tau^f \Delta t } \}$  
is than not only determined by the distribution $\rho(x_0)$ of $x_0$, but 
also by the probabilities of arriving along this precise route 
from $x_{-\tau^b \Delta t}$ in the past and
continuing  upto $x_{ +\tau^f \Delta t }$ in the future.
\begin{equation}
\label{equ:weight0} {\mathcal P}[{\bf x}]=\rho(x_0)
\prod_{i=-1}^{-\tau^b} p(x_{(i+1) \Delta t}\leftarrow x_{i\Delta t})
\prod_{i=1}^{\tau^f} p(x_{(i-1) \Delta t}\rightarrow x_{i\Delta t})
\end{equation}
where $p(x \rightarrow y)$ 
is the forward transition probability (more accurate: probability density) 
to go from $x$ to $y$ and 
$p(y \leftarrow x)$ is the probability that, if the system is at $y$, 
it came from $x$ in the past.
Here, $\rho(x_0)$ is not necessarily the Boltzmann distribution
or even a distribution in equilibrium. It is applicable
to all systems that (at least to some approximation)
are described by a steady state.
We can express $p(y \leftarrow x)$ in terms of forward transition 
probabilities as 
\begin{eqnarray}
\label{equ:ptilde}
p(y \leftarrow x)=\frac{\rho(x) p(x\rightarrow y)}{\int 
\mathrm{d} x' \rho(x') p(x' \rightarrow y)}=\frac{\rho(x) p(x\rightarrow y)}{\rho(y)}.
\label{Pback}
\end{eqnarray}
where $\int \mathrm{d} x'  \rho(x')  p(x'  \rightarrow y) = \rho(y)$
results from the steady state behavior.
Using this relation, one can show that 
Eq.~(\ref{equ:weight0}) is exactly equal to
the probability of the first point $\rho(x_{-\tau^b \Delta t})$ 
times the forward evolution probabilities
\begin{equation}
\label{equ:weight} {\mathcal P}[{\bf x}]=\rho(x_{-\tau^b \Delta t})
\prod_{i=-\tau^b}^{\tau^f-1} p(x_{i \Delta t}\rightarrow x_{(i+1)\Delta t})
\end{equation}
which is identical to the weight  
for a path in the TPS-ensemble \cite{Dellago02} with $x_{-\tau^b \Delta t}$
instead of $x_0$.
Note that so far, we have assumed nothing about the nature of dynamics 
(irreversible or reversible, stochastic or deterministic).
When restricted to the TIS  ensemble for interface $i$,
the probability density of a path can be written as 
\begin{equation}
\label{equ:TPE} {\mathcal P}_{\lambda_i}[{\bf x}]\equiv  \hat{h}_i({\bf x})  {\mathcal P}[{\bf x}] / Z(\lambda_i)
\end{equation}
where $\hat{h}_i$ is unity if the path goes from $\lambda_0$, crosses $\lambda_i$ 
and ends either at $\lambda_{i+i}$ or goes back to $\lambda_0$. 
Otherwise it is zero. The normalizing factor $Z(\lambda_i)$ equals
\begin{equation}
\label{equ:Z} Z(\lambda_i)\equiv\int\!\mathcal{D}{\bf x}\, \hat{h}_i({\bf x}) {\mathcal P}[{\bf x}]
\end{equation}
where the integral is taken over all possible paths ${\bf x} $ of all lengths, starting in all 
possible initial conditions $x_{-\tau_b}$.
Note that, contrary to TPS, Eq.~(\ref{equ:weight}) and  Eq.~(\ref{equ:TPE}) are not directly 
related to the relative probabilities of all paths in the TIS ensemble.
This is a result of the path ensemble containing paths of different lengths.
Eq.~(\ref{equ:weight}) turns into a true probability only when multiplied
with the infinitesimal volume element in path space $\mathcal{D}{\bf x} \equiv
\prod_{i=-\tau^b}^{\tau^f} \mathrm{d} x_{i \Delta t} \sim \mathrm{d} x^N$.
Hence, a long path has an infinitely smaller probability than a shorter one for stochastic 
dynamics.
Therefore, the concept of path space may sound peculiar for TIS.
Still, it is instrumental to derive proper acceptance rules for 
TIS obeying detailed balance.

When performing the random walk in the TIS path space using the shooting algorithm,
the detailed balance condition is
\begin{eqnarray}
\frac{P_\textrm{gen}[{\bf x}^{\rm (o)} \rightarrow {\bf x}^{\rm (n)}]}{ P_\textrm{gen}[{\bf x}^{\rm (n)} \rightarrow {\bf x}^{\rm (o)}]}
\frac{P_\textrm{acc}[{\bf x}^{\rm (o)} \rightarrow {\bf x}^{\rm (n)} ]}{ P_\textrm{acc}[{\bf x}^{\rm (n)} \rightarrow {\bf x}^{\rm (o)}]}=
\frac{{\mathcal P}_{\lambda_i}[{\bf x}^{\rm (n)}]}{{\mathcal P}_{\lambda_i}[{\bf x}^{\rm (o)}] }
\label{detbalpath}
\end{eqnarray}
where o and n, denote the  old and new path respectively.
The usual Metropolis acceptance rule is then
\begin{eqnarray}
P_{\rm acc}[{\bf x}^{\rm (o)} \rightarrow {\bf x}^{\rm (n)}]  = 
\hat{h}_i({\bf x}^{\rm (n)}) 
\min\left[ 1, \frac{{\mathcal P}_{}[{\bf x}^{\rm (n)}]}{{\mathcal P}_{}
[{\bf x}^{\rm (o)}]} \frac{P_{\rm gen}[{\bf x}^{\rm (n)} 
\rightarrow {\bf x}^{\rm (o)}]} 
{P_{\rm gen}[{\bf x}^{\rm (o)} \rightarrow {\bf x}^{\rm (n)}]}  \right],
\label{equ:acceptance}
\end{eqnarray}
Note that this rule only applies at the trajectory space level, it has nothing to do with whether the underlying dynamics is  stochastic or deterministic, or even reversible or irreversible.

The generation probability to create a new path from an old path using  the shooting move is 
given by
\begin{eqnarray}
{P_{\rm gen}[{\bf x}^{\rm (o)} \rightarrow {\bf x}^{\rm (n)}])]}& =& 
\frac{P(\delta p)}{N^{\rm (o)}}     
P_{\rm gen}^{\rm f} [{\bf x}^{\rm (o)} \rightarrow {\bf x}^{\rm (n)}]
P_{\rm gen}^{\rm b}[{\bf x}^{\rm (o)} \rightarrow {\bf x}^{\rm (n)}]
\label{Pgen}
\end{eqnarray}
where $ 1/N^{\rm (o)}$ is
the chance to choose the shooting point $\tau'$
with
$-\tau^{b{\rm (o)}} \leq \tau' \leq \tau^{f{\rm (o)}}$)
at the old path,
 $P( \delta p )$ the chance to select the randomized momenta displacements. 
As $\delta p$ is normally taken from a symmetric distribution, hence 
$P( \delta p )=P( -\delta p )$, this term will cancel in 
Eq.~(\ref{equ:acceptance}).
The last two factors in Eq.~(\ref{Pgen}) are the probabilities to  generate  
trajectories from the shooting point
point $\tau'$ with the new momenta.
These generation probabilities  are given by the underlying 
dynamics used to generate the trajectories. 
If one starts from a shooting point at $\tau^\prime$ on the old existing path  
(with $-\tau_b^{\rm (o)} < \tau^\prime < \tau_f^{\rm (o)}$) 
the generation probability 
for the forward segment is
\begin{equation}
\label{equ:shoot_gen_fw} P_{\rm gen}^{\rm f}[{\bf x}^{\rm (o)} \rightarrow {\bf x}^{\rm (n)}]
 =\prod_{i=\tau^\prime}^{\tau_f^{\rm (n)} -1} p\left(x_{i\Delta
t}^{\rm (n)} \rightarrow x_{(i+1)\Delta t}^{\rm (n)}\right).
\end{equation}
This generation probability is exactly identical to the weight of the forward  segment.
The integration of the backward segment is not always 
trivial especially when dealing with irreversible processes. 
However, in general, when reversible dynamics is applied,
the backward segment is obtained by reversing the momenta, 
integrating forward in time
and reversing the momenta again~\cite{Dellago02}.
Accordingly, 
the backward segment's  generation probability equals 
\begin{equation}
P_{\rm gen}^{\rm b}[{\bf x}^{\rm (o)} \rightarrow {\bf x}^{\rm (n)}]
=\prod_{i=-\tau_b^{\rm (n)}}^{\tau^\prime -1}  p \left( \bar x_{(i+1)\Delta t}^{\rm (n)}
\rightarrow \bar x_{i\Delta t}^{\rm (n)}\right).
\end{equation}
where $\bar x \equiv \{ r, -p\}$  for a phase point $ x \equiv \{ r, p\}$.
Using these generation probabilities and the path weight 
Eq.~(\ref{equ:weight}) the factor within the min function of 
Eq~(\ref{equ:acceptance}) can 
be written as
\begin{eqnarray}
\label{equ:ratio_back} 
\lefteqn{\frac{{\mathcal P}[{\bf x}^{\rm (n)}]
P_{\rm gen}[{\bf x}^{\rm (n)} \rightarrow {\bf x}^{\rm (o)}]}
{{\mathcal P}[{\bf x}^{\rm (o)}] P_{\rm gen}[{\bf x}^{\rm (o)} 
\rightarrow {\bf x}^{\rm (n)} ]}
= \frac{\rho [x_{-\tau_b^{\rm
\rm (n)}} ]N^{\rm (o)} }{\rho [x_{-\tau_b^{\rm (o)}}] N^{\rm (n)}}
\times}  \\
& & \prod_{i=-\tau_b^{\rm (n)}}^{\tau' - 1}\frac{p [x_{i\Delta t}^{\rm (n)}
\rightarrow x_{(i+1)\Delta t}^{\rm (n)}]}{
p[ \bar x_{(i+1)\Delta t}^{\rm (n)} \rightarrow \bar x_{i\Delta t}^{\rm
(n)}]}\prod_{i=-\tau_b^{\rm (o)}}^{\tau' - 1}\frac{ p [ \bar x_{(i+1)\Delta t}^{\rm (o)}
\rightarrow \bar x_{i\Delta t}^{\rm (o)}]}{p[x_{i\Delta t}^{\rm
(o)} \rightarrow x_{(i+1)\Delta t}^{\rm (o)}]} .\nonumber
\end{eqnarray}
where the generation probability and the weight of the forward parts of the 
trajectories have 
canceled each other. Eq.~(\ref{equ:weight}) simplifies tremendously
for dynamics that obey
the microscopic reversibility condition~\cite{Dellago02}
\begin{equation}
\label{equ:fw_bw}
\frac{p(x\rightarrow y)}{p(\bar y \rightarrow
\bar x)}=\frac{\rho_{\rm }(y)}{\rho_{\rm }(x)}.
\end{equation}
This condition can be seen as a special case of Eq.~(\ref{equ:ptilde}) for time-reversible dynamics with $p(y \leftarrow x) = p(\bar y \rightarrow
\bar x)$, but is very general and valid for a broad class of 
dynamics applying to both equilibrium and non-equilibrium systems. 
As result, 
if the microscopic reversibility condition~(\ref{equ:fw_bw})
is satisfied
and thus $p(\bar y \rightarrow \bar x)= p(y \leftarrow x)$,
almost all terms in Eq.~(\ref{equ:ratio_back})
cancel, except for the steady state distributions
$\rho(x_{\tau'})$ of the shooting points.
\begin{eqnarray}
P_{\rm acc}[{\bf x}^{\rm (o)} \rightarrow {\bf x}^{\rm (n)}]=
\hat{h}_i({\bf x}^{\rm (n)})  \min\left[1,\frac{\rho(x_{\tau'}^{\rm (n)})}{\rho
(x_{\tau'}^{\rm (o)})}  \frac{ N^{\rm (o)}}{N^{\rm (n)}}
\right].
\label{equ:acceptance_shoot_asymm}
\end{eqnarray}
which is exactly the same as for deterministic dynamics.

This is an important
result as it allows to perform the acceptance/rejection rule
(step 3 of Sec.~\ref{secshooting}) for the new momenta at the shooting
point even if the energy along the path changes giving a different weight to
$\rho(x_{\tau' \Delta t})$ as to $\rho(x_{0})$.
The ratio $N^{\rm (o)}/N^{\rm (n)}$ in
Eq.~(\ref{equ:acceptance_shoot_asymm}) can, of course, not be known in advance
at the shooting point. However, this is effectively circumvented
by defining $N_{\rm max}$ at step 4 of the
shooting algorithm in Sec.~\ref{secshooting}
leaving only the $\rho(x_{\tau'}^{\rm (n)})/\rho(x_{\tau'}^{\rm (o)})$
term for the acceptance rule (step 3).

\end{document}